\documentclass[aps,prd,notitlepage,reprint,twocolumn,showpacs,superscriptaddress,nofootinbib,floatfix]{revtex4-1}
\usepackage{amsmath, amsthm, amssymb,latexsym,mathrsfs,dsfont}    
\usepackage{graphicx}   
\usepackage{xcolor}
\usepackage{color}   
\usepackage[caption=false]{subfig} 
\usepackage{braket}
\usepackage{cancel}
\usepackage{array}
\usepackage{slashed}
\usepackage{bookmark,bbm}
\newcommand{\vect}[1]{\boldsymbol{#1}} 
\usepackage{cancel}
\graphicspath{{./Figs/}}
\newcommand{\eq}[1]{\begin{equation}\begin{aligned}#1\end{aligned}\end{equation}}
\newcommand{\ee}{\mathrm{e}}
\newcommand{\ii}{\mathrm{i}} 
\newcommand{\dd}{\mathrm{d}}

\DeclareMathOperator{\tr}{tr}

\newcommand{\lp}{{^{+}}}

\newcolumntype{P}[1]{>{\centering\arraybackslash}p{#1}}

\newcommand{\AN}[1]{\textcolor{black}{#1}}
\newcommand{\GI}[1]{\textcolor{black}{#1}}

\hypersetup{pdftitle={main.pdf},hidelinks
}
\begin{document}
	
	
	\title{Nucleon parton distributions from hadronic quantum fluctuations}

	\author{Andreas Ekstedt}
	
	\email{andreas.ekstedt@physics.uu.se}
	
	\affiliation{Department of Physics and Astronomy, Uppsala University, Box 516, SE-751 20 Uppsala, Sweden}

	\author{Hazhar Ghaderi}
	
	\email{hazhar.ghaderi@physics.uu.se}
	
	\affiliation{Department of Physics and Astronomy, Uppsala University, Box 516, SE-751 20 Uppsala, Sweden}

	\author{Gunnar Ingelman}
	
	\email{gunnar.ingelman@physics.uu.se}
	
	\affiliation{Department of Physics and Astronomy, Uppsala University, Box 516, SE-751 20 Uppsala, Sweden}
	
	\affiliation{Swedish Collegium for Advanced Study, Thunbergsv\"agen 2, SE-752 38 Uppsala, Sweden}

	\author{Stefan Leupold}
	
	\email{stefan.leupold@physics.uu.se}
	
	\affiliation{Department of Physics and Astronomy, Uppsala University, Box 516, SE-751 20 Uppsala, Sweden}
	
	\date{June 19, 2019 }
	
	\begin{abstract} 
		A physical model is presented for the non-perturbative parton distributions in the nucleon. This is based on quantum fluctuations of the nucleon into baryon-meson pairs convoluted with Gaussian momentum distributions of partons in hadrons. The hadronic fluctuations, here developed in terms of hadronic chiral perturbation theory, occur with high probability and generate sea quarks as well as dynamical effects also for valence quarks and gluons. The resulting parton momentum distributions $f(x,Q_0^2)$ at low momentum transfers are evolved with conventional DGLAP equations from perturbative QCD to larger scales. This provides parton density functions $f(x,Q^2)$ for the gluon and all quark flavors with only five physics-motivated parameters. By tuning these parameters, experimental data on deep inelastic structure functions can be reproduced and interpreted. The contribution to sea quarks from hadronic fluctuations explains the observed asymmetry between $\bar{u}$ and $\bar{d}$ in the proton. The strange-quark sea is strongly suppressed at low $Q^2$, as observed. 
		
	\end{abstract}
	
	
	
	\maketitle
	\section{Introduction}\label{sec:Introduction}
	The parton distribution functions (PDFs) of the nucleon are of great importance. One reason is that they provide insights into the structure of the proton and neutron as bound states of quarks and gluons, which is still a largely unsolved problem due to our limited understanding of strongly coupled QCD. Another reason is their use for calculations of cross-sections for high-energy collision processes. These factorize in a hard parton level scattering process, calculated in perturbation theory, and the flux of incoming partons given by the PDFs. 
	
	This involves the factorization of processes that occur at momentum-transfer scales of significantly different magnitudes. Of particular importance here is that the PDFs $f(x,Q^2)$ have the property that for $Q^2>Q_0^2\sim 1\,$GeV$^2$ the dependence on $Q^2$ can be calculated by the Dokshitzer-Gribov-Lipatov-Altarelli-Parisi equations (DGLAP) \cite{Altarelli:1977zs, Dokshitzer:1977sg, Gribov:1972ri} derived from perturbative QCD (pQCD), which is well-established theoretically and experimentally confirmed. However, the $x$-dependence needed at the starting scale $Q_0^2$ is not known from fundamental principles and instead parametrized to reproduce proton structure function data. This typically requires $x$-shapes given in terms of five parameters for each parton flavor, resulting in $\sim 30$ free parameters to account for valence quarks, gluons and sea quarks ($u$, $d$, $g$, $\bar{u}$, $\bar{d}$, $s$, $\bar{s}$). There are different collaborations \cite{Dulat:2015mca, Alekhin:2017kpj,Ball:2017nwa, Harland-Lang:2014zoa} performing such PDF parametrizations with DGLAP-based $Q^2$-evolution that give good fits of proton structure data and are excellent tools for cross-section calculations. However, the basic $x$-dependence at $Q_0^2$ originating from the bound-state proton is here only parametrized, but not understood.
	
	To understand the basic shape of the parton momentum distributions in physical terms, we here \GI{develop the theoretical basis for our earlier proposed model \cite{Edin:1998dz, Alwall:2004rd, Alwall:2005xd} and elaborate on its} phenomenologically successful results. The first basic idea is to use the uncertainty relation in position and momentum, $\Delta x\Delta p \sim \hbar/2$, to give the basic momentum scale of partons confined in the length scale $\Delta x$ given by the hadron diameter $D$. In the hadron rest frame it is natural to assume a spherically symmetric Gaussian momentum distribution with a typical width $\sigma \sim \hbar/(2D)$. The Gaussian \GI{is a convenient} mathematical form which cuts off large momenta that correspond to rare fluctuations. \GI{It may be motivated as resulting from many soft interactions within the hadron that add up to a Gaussian.} The strength of this approach lies in its simplicity and its small number of parameters. 
	
	The second basic idea is that whereas the valence quark and gluon distributions are essentially given by \GI{the basic description of the bound state bare nucleon and its quantum numbers,} the sea quark distributions are given by the hadronic fluctuations of the nucleon. For example, the proton quantum state $\ket{P} = \alpha_\text{bare}\ket{P}_\text{bare}  +\alpha_{P\pi^0}\ket{P\pi^0}+\alpha_{n\pi^+}\ket{n\pi^+} + \cdots$ contains not only the bare proton but also nucleon-pion fluctuations with probability amplitudes $\alpha_{N\pi}$. The point is that one should consider the dominant quantum fluctuations in terms of least energy fluctuation and thereby most long-lived \cite{Sullivan:1971kd,Ericson:1983um}. It is expected that pionic fluctuations dominate due to the small mass of the pion. In turn, its smallness compared to a typical hadronic scale $\sim 1\,$GeV is a consequence of spontaneous chiral symmetry breaking, which leads to the identification of pions as Goldstone bosons \cite{Peskin:257493, Scherer:2002tk}. From these dominant fluctuations \GI{of the proton state,} with the presence of $\pi^+$ but lack of $\pi^-$, one expects an asymmetry in the proton sea such that $\bar{d}>\bar{u}$ \GI{\cite{Thomas:1983fh,Melnitchouk:1991ui,Wakamatsu:1991yu,Henley:1990kw,Carvalho:1999he,Holtmann:1996be,Kumano:1997cy, Speth:1996pz, Kofler:2017uzq,Pobylitsa:1998tk,Dressler:1999zg}}, as is also observed in data \cite{Towell:2001nh}. \GI{In addition, an asymmetry in the $\bar{s}-s$ distributions is expected because the dominant $\Lambda K^{+}$ fluctuation gives a harder momentum distribution of the heavier $\Lambda$, and thereby of its $s$-quark, compared to the lighter $K^{+}$ and its $\bar{s}$ \cite{Alwall:2004rd, Alwall:2005xd, Signal:1987gz,Melnitchouk:1996fj,Melnitchouk:1999mv}.}
	
	\GI{
		This kind of hadronic fluctuations can in a simple phenomenological model \cite{Edin:1998dz, Alwall:2005xd} be handled by having the different baryon-meson ($BM$) fluctuation probabilities $|\alpha_{BM}|^2$ as free parameters fitted to data. Here, we instead follow the theoretically well-founded approach using the leading-order Lagrangian of three-flavor chiral perturbation theory \cite{Jenkins:1991es, Pascalutsa:1999zz, Pascalutsa:2006up, Ledwig:2014rfa} to describe the proton state as the Fock expansion
	}
	\eq{\label{E: FockExpansion}
		\ket{P} = \ &\alpha_\text{bare}\ket{P}_\text{bare}  
		+\alpha_{P\pi^0}\ket{P\pi^0}+\alpha_{n\pi^+}\ket{n\pi^+}
		\\
		+
		&
		\alpha_{\Delta^{++}\pi^-}\ket{\Delta^{++}\pi^-}
		+\alpha_{\Delta^{+}\pi^0}\ket{\Delta^{+}\pi^0}
		\\
		+
		&
		\alpha_{\Delta^{0}\pi^+}\ket{\Delta^{0}\pi^+}
		+\alpha_{\Lambda^0 K^+}\ket{\Lambda^0 K^+}
		+\cdots.
	}
	For a detailed account of the chiral symmetry basis for the baryon-meson Fock components, and subtle renormalization issues, we refer to \cite{Arndt:2001ye,Chen:2001eg,Chen:2001pva,Thomas:2000ny,Ji:2013bca,Wang:2016ndh}.
	The different terms in Eq.\ (\ref{E: FockExpansion}) are theoretically well defined and related to each other with only three coupling constants that are known from hadronic processes and weak decays of baryons. In addition to the probability for the different hadronic fluctuations, the theoretical formalism gives the hadron momentum distribution of the fluctuations. Incorporating the hadronic momentum distributions with the above parton momentum distributions in a hadron provides an improved model for the parton momenta of the proton quantum state.
	
	The PDFs are closely related to the proton structure functions that are measured in deep inelastic scattering (DIS) of leptons on protons. The most precise data are from electron and muon scattering, where the exchanged virtual photon has high resolution power and couples to quarks in the proton. The photon may therefore couple to a quark in the bare proton or in either the baryon or the meson in a baryon-meson fluctuation.

	\GI{
		Before giving the detailed account of model and results, we first elaborate on the model's general concepts and contrast it with other models. QCD covers various energy regimes where different methods apply. For the sector of light quarks, one has two energy scales at the microscopic level: the quark masses of a couple of MeV, and $\Lambda_{\rm QCD}$ of a couple of 100 MeV\textemdash above which the running coupling changes from strong to weak. Related to these microscopic scales one can on the observable hadronic side identify the low-energy regime on the order of the Goldstone boson (pion) masses (about 100 MeV), the medium-energy regime on the order of the typical hadron masses (about 1 GeV) and the high-energy regime much larger than the hadron masses. For the first and third of these regimes there exists systematic tools used to quantify the uncertainty of theoretical calculations and to improve on this uncertainty. The tools at our disposal are chiral perturbation theory for the low-energy regime and perturbative QCD for infrared safe quantities at high energies. Unfortunately, there are no systematic tools available at medium energies (except for lattice QCD \cite{Ma:2014jla,Orginos:2017kos,Lin:2014zya,Ji:2013dva}) to bridge the gap between these low- and high-energy regimes. Thus, for the medium-energy regime and for the non-perturbative quantities at high energies, one has to develop phenomenological models that preferably link as much as possible to the systematic approaches.}
	
	\GI{In our model, we make sure that the hadronic part matches to chiral perturbation theory at low energies. To extrapolate the PDFs from the medium-energy regime to higher energies we use the DGLAP-equations of  perturbative QCD. The essential new part is the explicit construction of the phenomenological model at a starting scale in the medium-energy regime. We assume, first, hadronic fluctuations play an important role because we operate in the medium-energy regime. Second, the quarks and gluons confined in a hadron experience so many soft interactions that their momentum distributions can be described by Gaussians. Clearly, these are assumptions as every phenomenological model is based on some assumptions. }
	
	\GI{Some other models do not have proper matching to the low-energy regime, in particular using a pseudoscalar instead of a pseudovector pion-nucleon coupling. The Goldstone theorem demands that all interactions of the pions vanish with the pion momenta, which is satisfied with the pseudovector interaction but not with the pseudoscalar interaction. Other approaches use sophisticated quark models to describe the quark-gluon aspects and/or interlink the hadronic and quark degrees of freedom, as e.g.\ in pion-quark models. In contrast to that, we clearly separate the hadronic fluctuations from the quark-gluon distributions inside of the hadrons, because these are described by different sets of quantum basis states with different degrees of freedom. Wherever proper QCD theory (governed by chiral perturbation theory or perturbative QCD) is not available,  we use simplicity as a useful guiding principle to get insights into the unknown dynamics of strongly interacting systems. The strength of our model lies in the clear links to the better known QCD regimes at higher and lower energies and otherwise in the simplicity of our model. We want to explore how far we can come with such a model and which insights can be obtained from it.
	}
	
	In this paper we present the complete model we have constructed based on these basic ideas.~Section \ref{Sec: DISonNucleonFluctuations} presents the formalism for DIS on the proton with its hadronic fluctuations, where some more technical details are provided in appendices at the end of the paper. In Section \ref{Sec: Distribution} we present our model for the parton distributions in a probed hadron, i.e.\ the $x$-shape at the starting scale $Q_0^2$ for pQCD evolution. Results are then presented in Section \ref{Sec: Results} in terms of obtained parton momentum distributions and their ability to reproduce data on proton structure functions and quark sea asymmetries. We give our conclusions in Section \ref{Sec: Conclusions}. 
	
	\section{DIS on a nucleon with hadron fluctuations}\label{Sec: DISonNucleonFluctuations}
	The cross-section for deep inelastic lepton-nucleon scattering  is theoretically well known as a product of the leptonic and hadronic tensors, $\dd \sigma \propto l^{\mu\nu} W_{\mu\nu}$. The leptonic tensor is straightforward to calculate and well known for photon exchange, $l^{\mu\nu} =\tr [\slashed{p}_{l}'\gamma^\mu \slashed{p}_l\gamma^\nu]/2$, as well as for $W$ or $Z$ exchange. We consider both electromagnetic and weak interactions. 
	
	\begin{figure}[!b]
		\centering
		\subfloat[]{\label{Fig: dis_bare}\includegraphics[width=0.33\columnwidth]{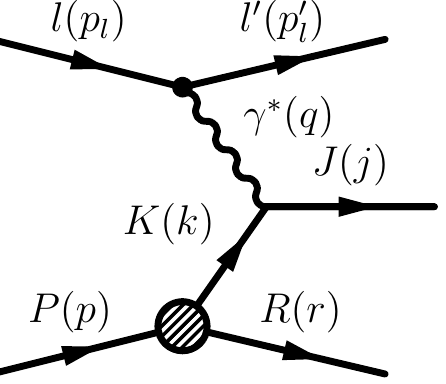}}
		\subfloat[]{\label{Fig: dis_meson}\includegraphics[width=0.33\columnwidth]{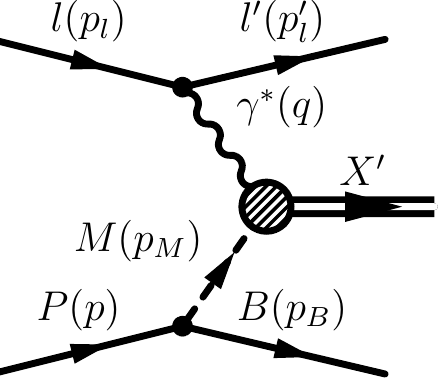}}
		\subfloat[]{\label{Fig: dis_baryon}\includegraphics[width=0.33\columnwidth]{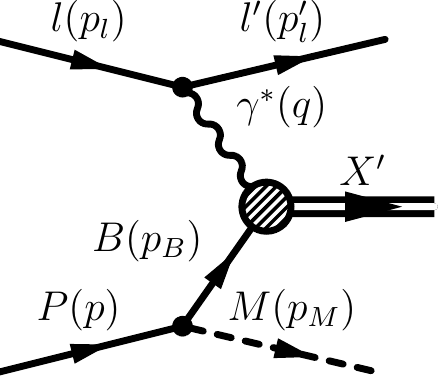}}
		\caption{Deep inelastic scattering on (a) the bare proton and on (b) the meson or (c) the baryon in a baryon-meson quantum fluctuation of the proton. }
		\label{Fig: DISfluc}
	\end{figure}
	The hadronic tensor $W_{\mu\nu}$ is a much more complex object and is of prime interest here. In order to take into account the proton target with its hadronic fluctuations, as illustrated in Fig.\ \ref{Fig: DISfluc}, we decompose the hadronic tensor to include the possibilities to probe either the bare proton or the meson or baryon in a fluctuation as follows 
	\eq{\label{E: HadTens2}
		W_{\mu\nu} = W_{\mu\nu}^{\text{bare}} + W_{\mu\nu}^{H} 
		=
		W_{\mu\nu}^{\text{bare}}+\sum_{BM}\left(W_{\mu\nu}^{MB}+W_{\mu\nu}^{BM}\right)
	} 
	where the notation $MB$ and $BM$ denotes probing the meson and baryon, respectively. 
	The general form of the hadronic tensor is \cite{Jaffe:161757} 
	\eq{
		W_{\mu\nu} = \frac{1}{4\pi} \int\! \dd^4\xi\, \ee^{\ii q \xi }\Braket{P|J_\mu(\xi)J_\nu(0)|P}
	}
	in terms of the hadronic current $J_\mu(\xi)$ as a function of the spacetime coordinate $\xi$. Using light-cone time-ordered perturbation theory \cite{Brodsky:1997de} we calculate the here introduced part corresponding to the hadronic fluctuations giving 
	\begin{widetext}
		\eq{\label{eq:HadTensor}
			W_{\mu\nu}^H = 
			\frac{1}{4\pi}\int\! \dd^4\xi\, \ee^{\ii q\xi}
			\sum_{BM,\lambda}
			\int_0^1\! \frac{\dd y}{y}\, 
			&
			\Big\{
			f_{MB}^{\lambda}(y) 
			\Braket{M(p_M^+=yp^+)|
				\left[J^{M}_\mu(\xi),J^{M}_\nu(0)\right]|
				M(p_M^+=yp^+)}
			\\
			\phantom{\sum_{BM,\lambda}
				\int_0^1\! \frac{\dd y}{y}\, 
				\Big\{
			}
			+
			&
			f_{BM}^{\lambda}(y) \Braket{B^{\lambda}(p_B^+=yp^+)
				|\left[J^{B}_\mu(\xi),J^{B}_\nu(0)\right]|B^{\lambda}(p_B^+=yp^+)}
			\Big\}  
		} 
	\end{widetext}
	where the first term is for DIS probing the meson ($M$) and the second term for probing the baryon ($B$). Expressions equivalent to (\ref{eq:HadTensor}) can be found in the literature \cite{Kofler:2017uzq}. 
	
	The integration variable is the fraction $y$ of the proton's energy-momentum carried by the meson or baryon. 
	Following common practice in DIS theory we use light-cone momenta $p^+=p^0+p^3$ and $p^-=p^0-p^3$, and thereby $y_i=\frac{p_i^{+}}{p^{+}}~(i=B,M)$. This has the advantage of being independent of longitudinal boosts, e.g.\ from the proton rest frame to the commonly used infinite-momentum frame. The light-cone momenta $p_i^{-}$ are given by the on-shell condition which in the $\vect{p}_\perp=0$ frame become 
	$p_i^{-}=\frac{m_i^2+k_\perp^2}{p_i^{+}},~(i=B,M)$.
	
	In (\ref{eq:HadTensor}) the sum runs over all baryon-meson pairs, with helicity $\lambda$ of the baryon. We have included all baryons in both the octet and decuplet of flavor SU(3), and all the Goldstone bosons represented by the mesons in the spin-zero octet. Naturally, the fluctuations with a pion will dominate due to its exceptionally low mass. Kaons are needed to get the leading contribution for the strange-quark sea. Table \ref{Table: Coupling} shows the relative strengths of different fluctuations due to the couplings to be discussed further below. 
	
	The dynamical behavior depends on the hadronic distribution functions 
	\eq{	
		f_{BM}(y) = \sum_\lambda f^\lambda_{BM}(y),
	}
	which are probability distributions for the physical proton to fluctuate to a baryon-meson pair. The baryon carries a light-cone fraction $y$ and the meson the remaining momentum fraction, i.e.\ satisfying the relation 
	\eq{
		f_{BM}(y)= f_{MB}(1-y), 
	}
	giving flavor and momentum conservation for each particular hadronic contribution. This ensures that all parton momentum sum rules come out correctly \cite{Szczurek:1992gs}. The hadronic distribution functions are explicitly given in Eqs.\ (\ref{E: HadDist},\ref{E: HadDist2}). Their explicit form depends on the Lagrangian used for the hadronic fluctuations, to which we now turn. 
	
	The relevant part of the leading-order chiral Lagrangian describing the interaction of spin 1/2 and spin 3/2 baryons with spin 0 mesons (as Goldstone bosons) is given by \cite{Jenkins:1991es, Pascalutsa:1999zz, Pascalutsa:2006up, Ledwig:2014rfa} 
	\begin{widetext}
		\eq{\label{E: DNpi}
			\mathcal{L}_\text{int} = \frac{D}{2}\tr(\bar{B} \gamma^\mu \gamma_5 \{u_\mu,B\})+\frac{F}{2}\tr (\bar{B}\gamma^\mu\gamma_5[u_\mu,B])
			-\frac{h_A}{m_R}\frac{\epsilon_{ade}g_{\mu\nu}
				\epsilon^{\rho\mu\alpha\beta}}{2\sqrt{2}}
			\left[\left(\partial_\alpha \bar{T}_\beta^{abc}\right)
			\gamma_5\gamma_\rho
			u^\nu_{bd} B_{ce}+\bar{B}_{ec}u^\nu_{db} 
			\gamma_5\gamma_\rho \partial_\alpha 
			T^{abc}_\beta\right], 
		}
	\end{widetext}
	where `tr' refers to flavor trace. Here, the $B_{ab}$ are the matrix elements of the matrix $B$ representing the octet baryons. The decuplet baryons are represented by the totally symmetric flavor tensor $T_{abc}^\mu$. Similarly, the spin 0 octet mesons are represented by a matrix $\Phi$ appearing in the Lagrangian through $u_\mu$ given by $u_\mu \equiv \ii u^\dagger(\nabla_\mu U)u^\dagger  = u_\mu^\dagger$ where $u^2 \equiv U = \exp(\ii \Phi /F_\pi)$. For further details see Appendix \ref{App: Lagrangian}. 
	
	From this Lagrangian we derive the non-zero terms when applied to our cases of a proton fluctuating into a meson together with an octet or decuplet baryon
	\begin{widetext}
		\eq{\label{E: octLag}
			\mathcal{L}^{P\to B_{\text{oct}}M} = &\Bigg[\frac{-D-F}{\sqrt{2} F_{\pi }}\bar{n} 
			\gamma^\mu\gamma_5 (\partial_\mu \pi^-)  
			-\frac{D+F}{2 F_{\pi }} \bar{P}
			\gamma^\mu\gamma_5 (\partial_\mu   \pi^0) 
			+
			\frac{D-3 F}{2\sqrt{3} F_{\pi }}\bar{P}
			\gamma^\mu\gamma_5 (\partial_\mu  \eta) 
			\\
			&-\frac{D-F}
			{2 F_{\pi }}\bar{\Sigma }^0 
			\gamma^\mu\gamma_5 (\partial_\mu K^-)
			-\frac{ D-F}
			{\sqrt{2} F_{\pi }}\bar{\Sigma}^+ 
			\gamma^\mu\gamma_5 (\partial_\mu  \bar{K}^0) 
			+  
			\frac{ D+3 F}{2\sqrt{3} F_{\pi }}\bar{\Lambda}
			\gamma^\mu\gamma_5 (\partial_\mu K^-)  
			\Bigg] P +\text{h.c.}
		}
		and 
		\eq{\label{E: decLag}
			\mathcal{L}^{P\to B_{\text{dec}} M} =\ &\frac{h_A 
				\varepsilon^{\rho\mu\alpha\beta}
			}{2m_R F_\pi}
			\Bigg[ 
			\sqrt{\frac{1}{3}}(\partial_\alpha \bar{\Sigma}^{*+}_\beta)
			\gamma_5\gamma_\rho(\partial_\mu \bar{K}^0)
			- \sqrt{\frac{1}{6}}(\partial_\alpha \bar{\Sigma}^{*0}_\beta)
			\gamma_5\gamma_\rho(\partial_\mu K^-)
			\\
			+
			&	 
			(\partial_\alpha \bar{\Delta}^{++}_\beta)
			\gamma_5\gamma_\rho(\partial_\mu\pi^+)
			- \sqrt{\frac{2}{3}}(\partial_\alpha \bar{\Delta}^{+}_\beta)
			\gamma_5\gamma_\rho(\partial_\mu\pi^0)
			- \sqrt{\frac{1}{3}}(\partial_\alpha \bar{\Delta}^{0}_\beta)
			\gamma_5\gamma_\rho(\partial_\mu\pi^-) 
			\Bigg]P +\!\text{h.c.}  
		}
	\end{widetext}
	respectively. 
	The effective nature of the hadronic theory ---manifested by the appearance of the derivative couplings of the form $\sim \gamma_5\gamma^\mu\partial_{\mu}M(z)$ in the Lagrangians---  introduces a slight ambiguity for the meson momentum $p_M$ appearing in the numerators in the application of the light-cone time-ordered framework. In the literature, there are two common choices for the meson momentum appearing in the numerators \cite{Holtmann:1996be, Kofler:2017uzq},
	\eq{\label{E: pionmomA}
		p_M^{(A)}=\left(p^{+}_P-p^{+}_B,p^{-}_P-p^{-}_B,\vect{p}_{P\perp}-\vect{p}_{B\perp}\right), 
	} 
	\eq{\label{E: pionmomB}
		p_M^{(B)}=\left(p^{+}_P-p^{+}_B,
		\frac{m_M^2+p_{M\perp}^2}{p_M^{+}},\vect{p}_{P\perp}-\vect{p}_{B\perp}\right). 
	}
	We find that these two choices give nearly identical results concerning the extracted values of our model's parameters and hence both choices yield similar conclusions. But even though choice (\ref{E: pionmomA}) gives a slightly better shape for the flavor asymmetry, to be discussed in Section \ref{Sec: ubardbarasymm}, we will use choice (\ref{E: pionmomB}) because this choice is in line with the Goldstone theorem \cite{PhysRev.127.965} whereas choice (\ref{E: pionmomA}) is not, as explicitly shown in Appendix \ref{Sec: VertexFunctions}.

	As discussed in Appendix \ref{App: Lagrangian}, the parameter values are as follows \cite{Granados:2017cib}. The pion decay constant $F_\pi = 92.4\text{ MeV}$ and the couplings $D=0.80,F=0.46$ \cite{Cabibbo:2003cu} and $h_A=2.7\pm 0.3$ with an uncertainty range to include partial decay width data on $\Delta\rightarrow N\pi$ and $\Sigma^{*}\rightarrow \Lambda\pi$ as well as the large-$N_C$ limit \cite{Dashen:1993as,Pascalutsa:2005nd} $h_A^\text{large-$N_C$} = \frac{3}{\sqrt{2}}g_A = 2.67$ where $g_A = F+D=1.26$ is well constrained by the beta decay of the neutron \cite{Patrignani:2016xqp}. 
	
	Using light-cone time-ordered perturbation theory, the Lagrangians (\ref{E: octLag},\ref{E: decLag}) lead to the hadronic distribution functions 
	\begin{widetext}
		\eq{
			\label{E: HadDist}
			&f^{\lambda}_{BM}(y)=\frac{1}	
			{ 2y(1-y)}\int\! \frac{\dd^2k_\perp}{(2\pi)^3} \,\left|
			g_{BM}~G(y,k_\perp^2,\Lambda^2)
			\frac{S^{\lambda}\left(y,\vect{k}	
				_\perp\right)}{m_P^2-m^2(y,k_\perp^2)}\right|^2,
		} 
		\eq{\label{E: HadDist2}
			&f^{\lambda}_{MB}(y)=\frac{1}
			{2y(1-y)}\int\! \frac{\dd^2k_\perp}{(2\pi)^3} \, \left|g_{BM}~
			G(1-y,k_\perp^2,\Lambda^2)\frac{S^{\lambda}\left(1-y,\vect{k}_\perp\right)}{m_P^2-m
				^2(1-y,k_\perp^2)}\right|^2
		}
	\end{widetext}
	for the baryon and meson, respectively, probed in the fluctuation. As required, they satisfy $f_{BM}(y) = f_{MB}(1-y)$. The various hadronic couplings $g_{BM}$ are provided in Table \ref{Table: Coupling} and the vertex functions $S^{\lambda}(y,\vect{k}_\perp)$ are given in Appendix \ref{Sec: VertexFunctions}. The suppression of the energy fluctuation is seen as the propagator with the difference of the squared masses of the proton and the baryon-meson system given by 
	\eq{\label{E: MesonBaryonMass}
		m^2(y,k_\perp^2)\equiv\frac{m_B^2+k_\perp^2}{y}+\frac{m_M^2+k_\perp^2}{1-y}. 
	}  
	\begin{table*}\caption{The proton to baryon-meson fluctuations, with couplings and their strength relative to the respective largest coupling $g_{BM}^\text{max}$, where $g_{DM}^\text{max}=g_{\Delta^{++}\pi^-}$ and $g_{OM}^\text{max} = g_{n\pi^+}$. }\label{Table: Coupling}
		\renewcommand{\arraystretch}{1.5}
		\begin{tabular}{ | c || c | c | c | c | c || c | c | c | c | c | c |}
			\hline
			$BM$ &  $\Delta^{++}\pi^-$ & $\Delta^{+}\pi^0$ & $\Delta^{0}\pi^+$ 
			& $\Sigma^{*+}K^0$ & $\Sigma^{*0}K^+$ 
			& $n\pi^+$  & $P\pi^0$ & $\Lambda K^+ $ & $\Sigma^+K^0$
			& $\Sigma^{0}K^+$ & $P\eta$ 
			\\
			\hline
			$g_{BM}$ & $\frac{h_A}{2m_R F_\pi}$ & $\frac{-h_A}{\sqrt{6}m_R F_\pi}$ 
			& $\frac{-h_A}{2\sqrt{3}m_R F_\pi}$ & $\frac{h_A}{2\sqrt{3}m_R F_\pi}$ & 
			$\frac{-h_A}{2\sqrt{6}m_R F_\pi}$ 
			& $\frac{-D-F}{\sqrt{2}F_\pi}$ & $\frac{-D-F}{2F_\pi}$ & 
			$\frac{D+3F}{2\sqrt{3}F_\pi}$ & $\frac{-D+F}{\sqrt{2}F_\pi}$ 
			& $\frac{-D+F}{2F_\pi}$ & $\frac{D-3F}{2\sqrt{3}F_\pi}$ 
			\\
			
			$\left|\frac{g_{BM}}{g_{BM}^\text{max}}\right|^2 $ & 1 & 0.67 & 0.33 & 0.33 & 0.17
			& 
			1 & 0.5 & 0.5 & 0.08 & 0.04 & 0.03   
			\\
			\hline
		\end{tabular}
	\end{table*}
	
	The function $G(y, k_\perp^2,\Lambda_H^2)$ is a cut-off form factor, which is used to avoid the integral getting an unphysical divergence. The physics issue to account for is the fact that the description in terms of hadronic degrees of freedom is only valid at hadronic scales, whereas for higher momentum-transfer scales parton degrees of freedom should be used. To phase out the hadron formalism it is convenient to introduce a suitably constructed form factor. 
	
	In practice, it is conceivable to cut on the virtuality of the fluctuation \cite{Holtmann:1996be} or on the modulus of the three-momentum (in a proper reference frame). While the first option sounds plausible from a point of view of Heisenberg's uncertainty relation (or Fermi's Golden Rule), this quantum-mechanical aspect is already accounted for by the just mentioned propagator in Eqs.\ (\ref{E: HadDist},\ref{E: HadDist2}). An additional such cut is therefore artificial. Instead we choose to cut off the three-momentum of the hadrons in the fluctuation as seen in the rest frame of the proton. If relevant at all, high-momenta fluctuations should be of partonic not hadronic nature. 
	
	To conserve the condition $f_{BM}(y)= f_{MB}(1-y)$ it is necessary to use a symmetric combination of the meson/baryon three-momentum and a natural choice is to use the average of the squares of the three-momenta of the meson and baryon. To make this manifestly frame-independent we write its value in the proton rest frame expressed in a Lorentz-invariant form and take the form factor to be 
	\eq{
		G\left(y,k_\perp^2,\Lambda_H^2\right)
		=\exp\left[-\frac{\mathcal{A}^2\left(y,k_\perp\right)}
		{2\Lambda_H^2}\right] 
	}
	where $\Lambda_H$ is the parameter that regulates the suppression of larger scales. Because this is related to the switch to partonic degrees of freedom, one would expect it to be of the same order as the starting scale $Q_0$ of the pQCD formalism. The function $\mathcal{A}^2$ in the form factor is given by 
	\begin{widetext}
		\eq{\label{E: FFmomentum}
			&\mathcal{A}^2\left(y,k_\perp\right)\equiv
			\left(\vect{p}^{2}_{B}
			+\vect{p}^{2}_{M}\right)|_{p^{+}=m_P}
			=
			\frac{(p\cdot p_B)^2+(p\cdot p_M)^2}{m_P^2}-m_B^2-
			m_M^2
			\\
			&= 
			\left(
			\frac{m_B^2+k_\perp^2}{2 m_P y}
			\right)^2
			+
			\left(\frac{m_M^2+k_\perp^2}{2 m_P (1-y)}\right)^2
			+k_\perp^2-\frac{m_B^2+m_M^2}{2}
			+\frac{m_P^2}{4}\left[(1-y)^2+y^2\right]
			, 
		}
	\end{widetext}
	where light-cone momenta $p_B^+ = y p^+$ and $p_M^+= (1-y)p^+$ have been used to obtain the last expression. This form factor regularizes any potential end-point ($y=0,1$) singularities. Furthermore, high values of $k_\perp$ are largely suppressed which renders the integrals in Eqs.\ (\ref{E: HadDist},\ref{E: HadDist2}) finite and restricts the hadronic fluctuations to the low-momentum scales where the hadronic language is applicable. 
	
\begin{figure}[t]
		\centering
		\includegraphics[width=1\linewidth]{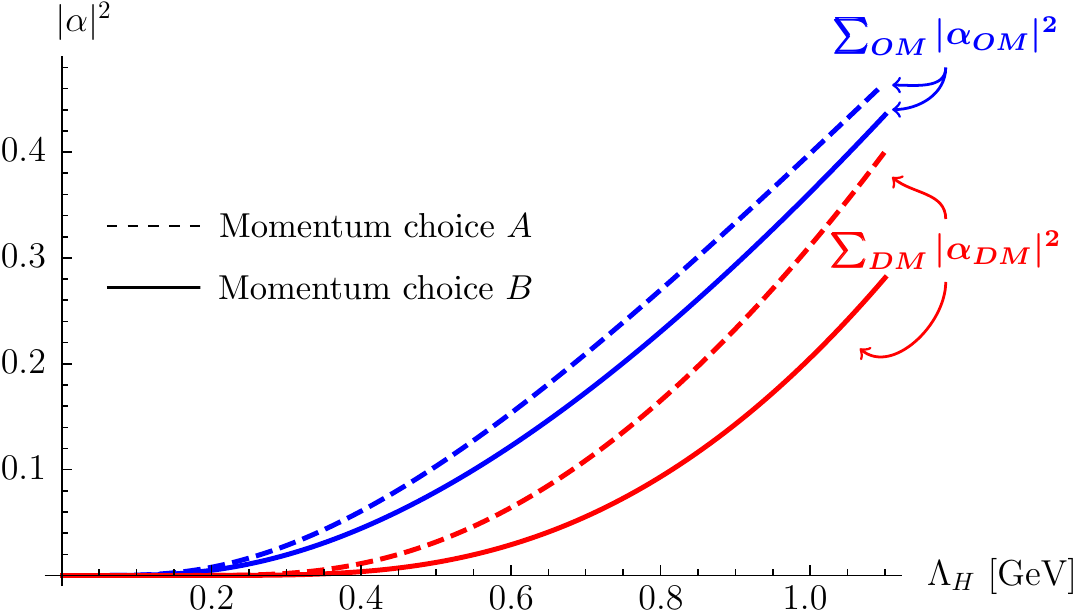}
		\caption{\label{Fig: CompareProbABappendix}The baryon-meson fluctuation probability as a function of the cut-off parameter $\Lambda_H$ for two different choices of meson momentum. The solid [dashed] curves refer to choice (\ref{E: pionmomB}) [(\ref{E: pionmomA})]. The two upper curves (blue) are the sum of all the octet-baryon--meson probabilities. The two lower curves (red) are the sum of all the decuplet-baryon--meson probabilities. }
	\end{figure}

	Using this theoretical formalism we illustrate the total fluctuation probability for a proton to a $BM$ pair by calculating 
	\eq{
		\lvert \alpha_{BM}(\Lambda_H)\rvert^2  = \int_0^1\! \dd y\, f_{BM}(y)
	}
	for both momentum choices, Eqs.\ (\ref{E: pionmomA},\ref{E: pionmomB}), 
	giving the result shown in Fig.\ \ref{Fig: CompareProbABappendix}. One observes that the probability for a proton to fluctuate into a baryon-meson state is quite sizable. Notably, for a cut-off $\Lambda_H$ around $1\,$GeV, the contribution from the baryon-decuplet members (mainly from the $\Delta$'s) is comparable in size to the nucleon-pion fluctuations.
	
	Due to the hadronic fluctuations, the PDFs for the proton are given by a convolution of the hadronic distributions, Eqs.\ (\ref{E: HadDist},\ref{E: HadDist2}), and the PDFs for the hadron being probed. Thus, the PDF for a parton $i$ in the proton can be written in the form \cite{Holtmann:1996be, Pasquini:2006dv, Kofler:2017uzq} 
	\eq{
		f_{i/P}&(x)= ~ f_{i/P}^{\text{bare}}(x)
		\\
		+& \sum_{H\in B,M}\int\!  \dd y\, \dd z \,\delta(x-y z)f^{\text{bare}}_{i/H}(z) f_{H/P}(y), 
	}
	taking into account the contributions from the bare proton and the $BM$ fluctuations. In our approach, the PDF for the `bare' part in any of these contributions (bare proton, baryon or meson in a fluctuation) is obtained from a Gaussian as mentioned in the Introduction and to be discussed in Section \ref{Sec: Distribution}. These bare distributions contain constituent quarks and gluons, but no sea quarks.
	
	In this work we include all the admissible octet-baryon--meson and decuplet-baryon--meson pairs in the fluctuations, i.e.\ the $\ket{N\pi}$, $\ket{\Delta\pi}$, $\ket{\Lambda K}$, $\ket{\Sigma K}$, and $\ket{\Sigma^{*} K}$ fluctuations. The $\ket{P\eta}$ contribution can be neglected due to mass suppression and its very small coupling to the proton state, see Table\ \ref{Table: Coupling}. The nucleon-pion and the Delta-pion fluctuations give the largest contributions, while the $\ket{\Lambda K}$, $\ket{\Sigma K}$ and $\ket{\Sigma^{*}K}$ fluctuations act as small corrections. However, because neither $\ket{N\pi}$ nor $\ket{\Delta\pi}$ contribute to the strange sea, other fluctuations like $\ket{\Lambda K}$ while being small are the leading hadronic contributions to the strange sea. It is found that the $\ket{\Lambda K}$ fluctuation is most important, while the $\ket{\Sigma K}$ and $\ket{\Sigma^*K}$ fluctuations are suppressed due to a small coupling and the larger masses involved respectively, see Table \ref{Table: Coupling}. 
	
	Once the starting distributions have been obtained from the convolution model at a particular starting scale $Q^2_{0}$, the PDFs are obtained for higher $Q^2$ by DGLAP evolution. The DGLAP evolution is performed at next-to-leading order (NLO) using the QCDNUM package \cite{Botje:2010ay}. 
	
	\section{Generic model for parton distributions in a hadron}\label{Sec: Distribution} 
	
	For any bare hadron we are considering the parton momentum distributions for its valence quarks/antiquarks and a gluon component. This applies for both the above considered bare proton as well as for the baryons and mesons in a hadronic fluctuation. 
	\GI{
		We therefore now consider deep inelastic scattering (DIS) on such a generic hadron. The DIS formalism was developed and is conventionally interpreted in the infinite momentum frame (IMF) where the hadron has a large momentum such that parton masses and transverse momenta are kinematically negligible. The essential momentum axis is defined through the  measurement given by the probe, e.g. a virtual photon. The formalism is \cite{Zavada:1996kp} also applicable in the target hadron rest frame and does not prefer any special reference system, with IMF as a limiting case. Our model for PDFs of the bound hadron state has a basic physics motivation originating in the hadron rest frame, but because we define the parton's energy-momentum fraction $x={k^+}/{p_H^+}$ using light-cone momenta ($k^+$ for the parton and $p_H^+$ for the hadron) this key variable is independent of longitudinal boosts. Thereby our model should be applicable in any frame of interest for DIS. 
	}
	
	In the rest frame of such a generic bare hadron there is no preferred direction. Therefore the spherical symmetry motivates the assumption that the parton's momentum distributions in $k_x, k_y$ and $k_z$ are the same. Assuming a Gaussian momentum distribution for these components provides a convenient mathematical form which suppresses large momenta that should correspond to rare momentum fluctuations. 
	\GI{
		Because a Gaussian results mathematically from adding many small contributions, it can be argued to be applicable here to represent the added effect of many soft momentum exchanges within the bound-state hadron in the absence of a proper description derived from QCD. The four-momentum distribution for a parton of type $i$ and mass $m_i$ is therefore assumed to be given by \cite{Edin:1998dz, Alwall:2005xd} 
	}

	\eq{\label{E: F-bare-k}
		F_{i/H}(k) 
		&=
		N_{i/H}(\sigma_i,m_i)
		\\
		\times
		&\exp\left(-\frac{(k_0-m_i)^2+k_x^2+k_y^2+k_z^2}{2\sigma_i^2}\right) 
	}
	where $N$ is a normalization factor.
	
	The width $\sigma$ of this Gaussian is expected to be physically given by the uncertainty relation $\Delta x\Delta p \sim \hbar/2$ that enforces increasing momentum fluctuations for a particle confined in a smaller spatial range. Thus, for a hadron of size $D$ (diameter) one expects $\sigma \sim \hbar/(2D)$ and therefore being typically of order $0.1\,$GeV. 
	
	\GI{
		The PDF for a parton $i=q$, $\bar{q}$, $g$ of mass $m_i$ in the hadron $H$ is then given by 
		\eq{\label{E: f-bare-x}
			f^\text{bare}_{i/H}(x) = ~&\int' \!\frac{\dd^4k}{(2\pi)^4}\, 
			\delta\left(\frac{k\lp}{p_H^+}-x\right)F_{i/H}(k) 
		}
		where $x={k^+}/{p_H^+}$ is the discussed light-cone energy-momentum fraction of a parton in the hadron. 
		The prime on the integral sign in (\ref{E: f-bare-x}) indicates the kinematical constraints that the quark four-momentum $k$ must obey. We demand
		that the scattered parton must be on-shell or have a timelike virtuality (causing final-state QCD radiation) limited by the
		mass of the hadronic system, i.e.\ $m_i^2 < j^2 = (k+q)^2 < (p_H+q)^2$. Likewise, the hadron remnant must have a
		four-vector $r^2=(p_H-k)^2>0$, cf.\ Fig.\ \ref{Fig: dis_bare}.
	}
	
	\GI{
		From a conceptual point of view it is interesting to note that
		the constraints involving the photon momentum $q$ bring in the influence of the quantum mechanical measurement process on the
		distribution. Thus the originally spherically symmetric function (\ref{E: F-bare-k}) is reshaped into a distribution that
		contains the directional information originating from the virtual photon that probes the hadron. 
	}
	
	\GI{
		As one consequence of these kinematical constraints, the light-cone energy-momentum fraction $x$ is automatically restricted
		to its physical range $0<x<1$.
		The constraint that the remnant should have timelike momentum, $(p_H-k)^2 > 0$, has a related
		interesting consequence. In the hadron rest frame this constraint translates to
		\begin{eqnarray}
		k_\perp^2 < x (1-x)m_H^2 - (1-x) k^2  \,.
		\label{eq:ineq1}
		\end{eqnarray}
		Obviously this inequality is then true for any frame that leaves $k_\perp$ untouched, i.e.\ any frame between the hadron rest
		frame and the infinite-momentum frame. On account of (\ref{eq:ineq1}), the integral measure $k_\perp \, \dd k_\perp$, which is a
		part of $\dd^4 k$ in (\ref{E: f-bare-x}), vanishes for $x \to 1$. Thus, in this limit the PDF smoothly vanishes.
		To summarize, the kinematical
		constraints ensure that $f^\text{bare}_{i/H}(x)$ is only non-vanishing for $0<x<1$
		and that $\lim\limits_{x \to 1} f^\text{bare}_{i/H}(x) = 0$.
	}
	
	The normalizations $N_{q/H}(\sigma_q,m_q)$ are fixed by the flavor sum rules, i.e.\ the integrals giving the correct numbers of different valence quark flavors. $N_{g/H}(\sigma_g,0)$ is fixed by the momentum sum rule, i.e.\ to get the sum of $x$-weighted integrals to be unity.
	
	Thus, the only free parameters are the Gaussian widths $\sigma_g$, $\sigma_1$, $\sigma_2$, where the indices refer to the widths of the distributions for the gluon and for the quark flavors represented by one quark ($\int_0^1\! \dd x\, f^\text{bare}_{q/H}(x) =1$) or two quarks ($\int_0^1\! \dd x \, f^\text{bare}_{q/H}(x)=2$) in the probed hadron. For instance, $\sigma_2$ ($\sigma_1$) applies for $u$ ($d$) in the proton. 
	\GI{
		Distributions of quarks appearing triply in a baryon, such as the $u$ distribution in the $\Delta^{++}$ baryon, could be given a different Gaussian width $\sigma_3$. However, the final distributions are not very sensitive to $\sigma_3$. For simplicity we choose to determine such distributions by making use of isospin symmetry relations such as $f^\text{bare}_{u/\Delta^{++}}(x)=2 f^\text{bare}_{u/\Delta^{+}}(x)-f^\text{bare}_{d/\Delta^{+}} (x)$.
	}
	
	With this model we have chosen a minimalistic approach with the same Gaussian distributions for all partons, having a width that only depends on the number of same-flavor quarks, but not on the particular hadron considered. Of course, one could introduce more complexity requiring more parameters, but we find it more interesting to see what insights this minimal physics-motivated model can give. 
	
	The above parametrization automatically conserves isospin (e.g.\ $f^\text{bare}_{u/P}(x)=f^\text{bare}_{d/n}(x)$ and similarly for the other hadrons). With the above widths for all possible hadrons, the distributions only depend on mass effects via the mentioned kinematical constraints. 
	
	It should be noted that these PDFs can be analytically evaluated in terms of error functions \cite{Alwall:2005xd}, but in practice it is more convenient to evaluate them numerically. As discussed, these bare distributions will only contain valence quarks and gluons, whereas the sea distributions will be entirely generated by hadronic fluctuations. All the resulting PDFs are at the low hadronic scale to be used as starting distributions at $Q_0^2$ for DGLAP evolution to large scales $Q^2$. 
	
	\section{Model results based on data comparison}\label{Sec: Results}
	\subsection{The few adjustable parameters}\label{Sec: parameters}
	
	The model introduced above has few parameters which are expected to lie in a limited range in order for the model to make sense. 
	
	The description of hadronic fluctuations is controlled by three coupling strengths with values already fixed by data from various hadronic processes. As discussed in connection with Table \ref{Table: Coupling} above, the coupling $g_A=F+D=1.26$ is constrained to the $1\%$ level from the beta decay of the neutron \cite{Patrignani:2016xqp} 
	whereas $D=0.80$ and $F=0.46$ may vary independently by $\sim \pm 5\%$ as long as their sum is fixed \cite{Cabibbo:2003cu}. Because it is their sum that appears in the most probable fluctuations, a variation in $D$ and $F$ has a negligible effect on the results. For the decuplet coupling we take $h_A=2.7 \pm 0.3$. Because $h_A/m_R$, with $m_R$ the resonance mass (basically $m_\Delta$), appears as the effective coupling in the decuplet Lagrangian (\ref{E: decLag}), we vary the ratio 
	\eq{\label{E: 1020}
		1.737 ~\text{GeV}^{-1} < h_A/m_\Delta < 2.435 ~\text{GeV}^{-1}
	} 
	to see the resulting sensitivity on this uncertainty (see Appendix \ref{App: Lagrangian} for details).  
	
	The only newly introduced parameter in the hadron fluctuation model is the regulator for the high-momentum suppression, $\Lambda_H$. This parameter is constrained to have a value large enough to allow hadronic fluctuations of some baryon-meson configurations, i.e.\ energy fluctuations of at least a few hundred MeV. On the other hand, it must be small enough to ensure a separation between the hadronic and partonic degrees of freedom. Thus, a reasonable expectation is a value in the range $0.5 ~\text{GeV}\lesssim \Lambda_H\lesssim 1$ GeV.
	
	Based on the model it is expected that the  $\sigma$ parameters have values on the order of $0.1\,$GeV and the $Q_0$ a value on the order of $1\,$GeV.  
	The former is given, as discussed above, by the inverse size of hadrons and the latter by the factorization scale of non-perturbative bound hadron state dynamics from the pQCD description
	\GI{
		at higher momentum scales. For the evolution of parton density functions at $Q^2>Q_0^2$ we use the NLO DGLAP equations with the running coupling $\alpha_s(Q^2)$, with $\alpha_s(m_Z^2)=0.1190$ in agreement with the measured value \cite{Tanabashi:2018oca}. In addition, we expect that $Q_0 \sim \Lambda_H$. However,
	}because $Q_0$ and $\Lambda_H$ are defined in two different formalisms, the partonic and hadronic respectively, and there is no theoretically well-defined link between these two descriptions, one cannot {\it a priori} take them as being the same parameter. Still, as will be seen below they do come out to have the same value within their uncertainties.
	
	\subsection{Comparison with proton structure function data}\label{sec:StructureFunctions}
\begin{figure}	
	\centering	
			\subfloat[]{\label{Fig: F2Large}
			\scalebox{1.0}{
	\includegraphics[width=1\columnwidth,trim=0cm 0.8cm -0.2cm 0cm,clip]{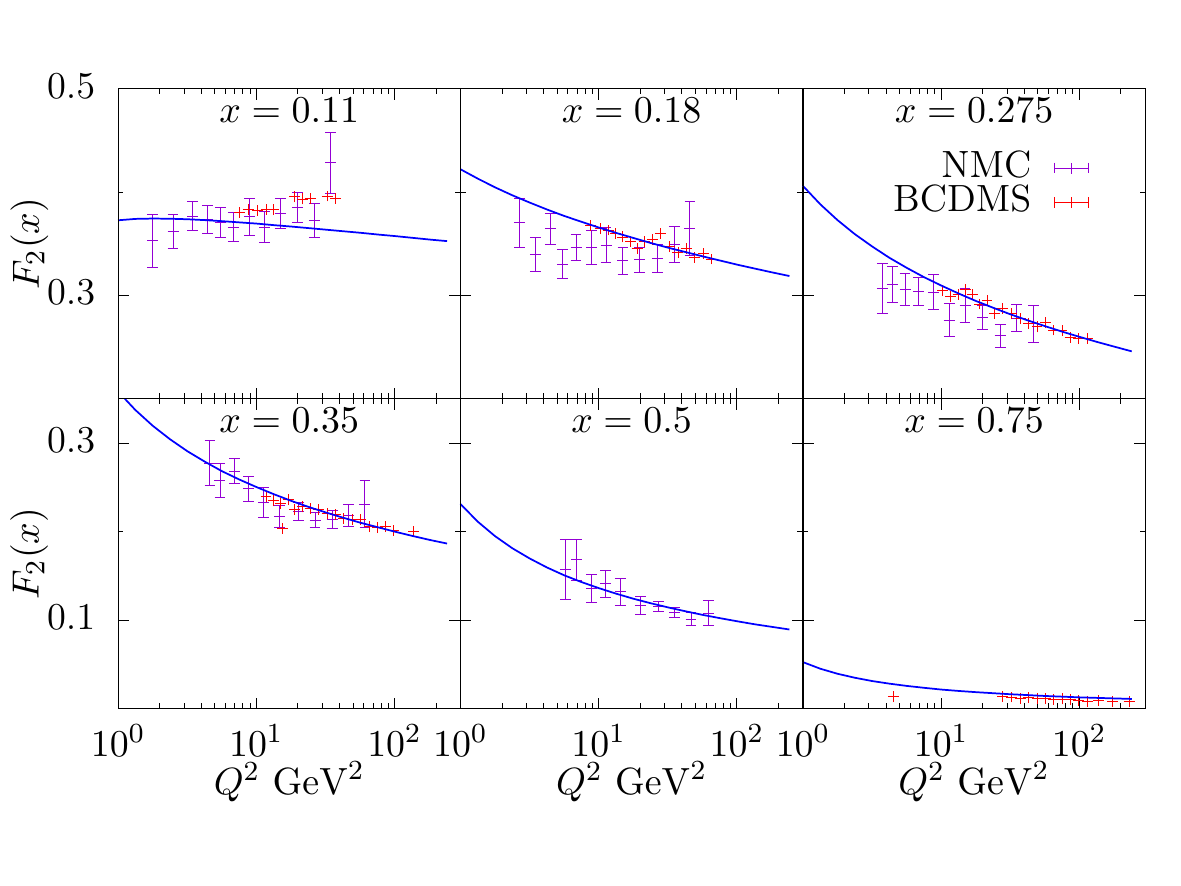}
	}
	}
	\par
		\subfloat[]{\label{Fig: F3nutevCDHSW}
		\scalebox{1.0}{	
	\includegraphics[width=1\columnwidth,trim=0cm 0.1cm 0cm 0cm,clip]{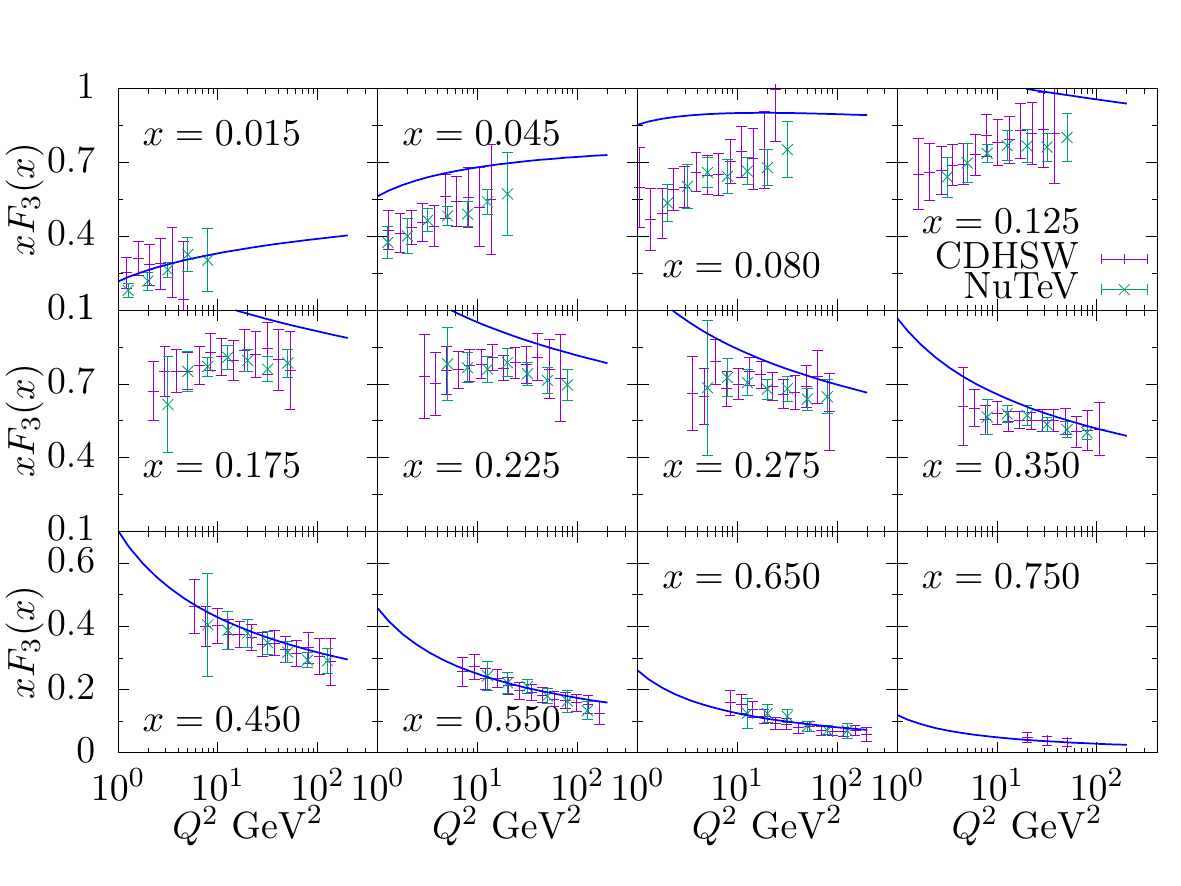}
			}
		}\par
		\subfloat[]{\label{Fig: F2neu}\scalebox{1.0}
		{			\includegraphics[width=1\columnwidth,trim=0cm 0.2cm 0cm 0.50cm,clip]{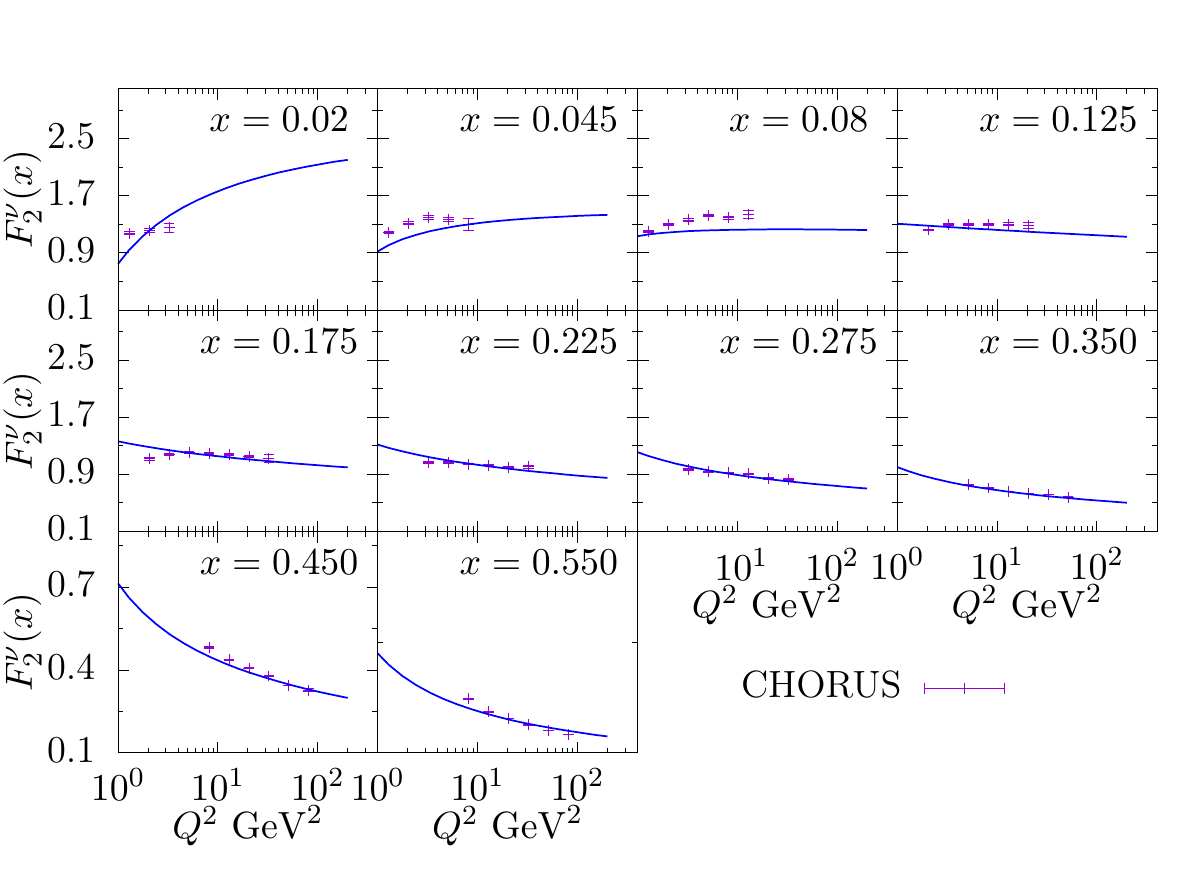}
			}
		} 
		\caption{\label{Fig: F3F2} Model predictions for proton structure functions \protect\subref{Fig: F2Large} $F_2$, \protect\subref{Fig: F3nutevCDHSW} $xF_3$, and \protect\subref{Fig: F2neu} $F_2^\nu$ as function of $Q^2$ for different $x$-bins compared
		 to data on fixed target $\mu P$ scattering from the New Muon Collaboration (NMC) \cite{Arneodo:1995cq} and BCDMS \cite{Benvenuti:1989rh}, and from neutrino-scattering experiments CDHS \cite{Berge:1989hr}, NuTeV \cite{Tzanov:2005kr}, and CHORUS \cite{Onengut:2005kv}.}
	\end{figure}

\begin{figure*}
		\scalebox{1.1}{
			\includegraphics[width=1\textwidth,trim=1.2cm 1.0cm -0.5cm 0cm,clip]{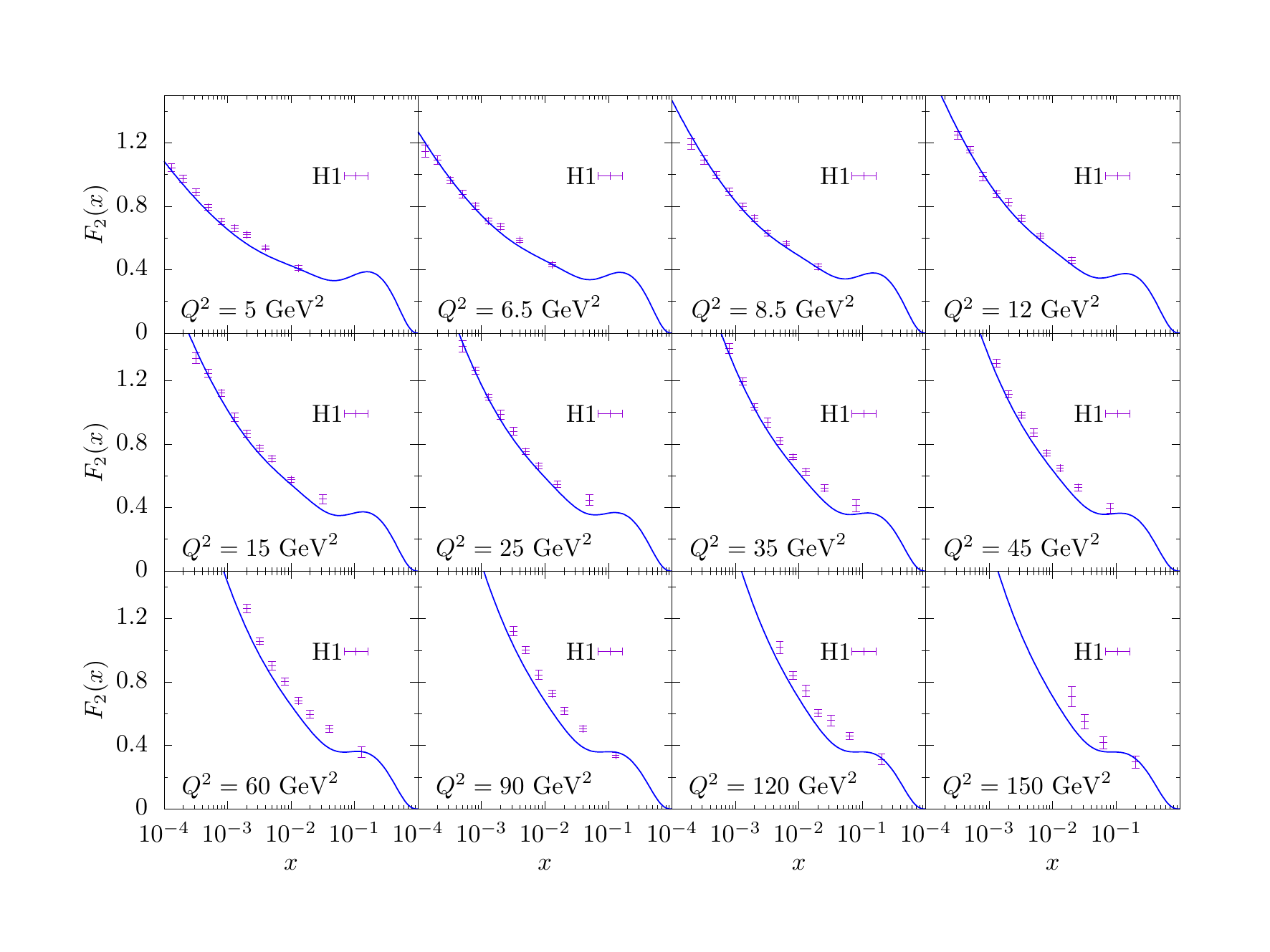}
		}
		\caption{The proton structure function $F_2$ as a function of $x$ for various $Q^2$-bins. Our model curve compared to data from the H1 $eP$ collider experiment \cite{Adloff:2000qk}.}
		\label{Fig: F2smallx} 
	\end{figure*}

	The values of the just discussed parameters are obtained from inclusive deep inelastic lepton-proton scattering giving the proton structure functions $F_2$ and $xF_3$. Figures \ref{Fig: F3F2}-\ref{Fig: F2smallx} show $\mu P$ data from NMC \cite{Arneodo:1995cq} and BCDMS \cite{Benvenuti:1989rh},  
	neutrino data from CDHSW, NuTeV  and CHORUS \cite{Berge:1989hr,Tzanov:2005kr,Onengut:2005kv} and $eP$ data from H1 \cite{Adloff:2000qk} in comparison to our model results.
	\GI{
		Our objective is not to obtain the best possible fit in terms of lowest $\chi^2$ in a global fit of all relevant data and thereby compete with conventional PDF parametrizations of the different $xf_i(x,Q_0^2)$ having some 30 free parameters. Instead, our aim is to gain understanding through our physically motivated model with only 5 parameters of physical significance and with expected values in order for the model to make sense. We have therefore not made a global fit to all data, but rather investigated the importance of our few parameters for different observables.
	}
	The parameters $\sigma_1$, $\sigma_2$, $\Lambda_H$, and $ Q_0$ can be nailed down using $F_2$ and $xF_3$ data. $Q_0$ and $\sigma_g$ are given by the small-$x$ $F_2$ data: With $Q_0$ given, it's always possible to fit data by varying $\sigma_g$. We find that the following parameter values give the best overall result 
	\eq{\label{E: Parametervalues}
		&\sigma_1=0.11~\text{GeV},
		\sigma_2=0.22~\text{GeV},
		\sigma_g=0.028~\text{GeV},\\ 
		&\Lambda_H=0.87~\text{GeV}, 
		Q_{0} = 0.88~\text{GeV}. 
	}
	
	Notice that the fit results in $\Lambda_H$ and $Q_0$ being practically the same, confirming our expectation that this scale constitutes the transition from hadron to parton degrees of freedom in the model. Moreover, the Gaussian widths are found to be of the expected magnitude $\sim 0.1$~GeV. The gluon distribution is particularly soft, which may seem surprising. However, the above argument based on the uncertainty relation gives $\sigma \sim \hbar/(2D)= 56$~MeV for the proton charge radius 0.875~fm \cite{Patrignani:2016xqp}.  
	In view of the symmetry properties of two-particle wave functions of indistinguishable states it should not be surprising that the momentum distribution for quark flavors that appear singly in the hadron differ from the one for quark flavors that appear pairwise.

	Considering the fact that the model has effectively only four parameters, which are also constrained by the physics assumptions of the model, it is remarkable that such a large amount of structure function data can be reasonably well described. Admittedly, there are some kinematical regions of some experimental data sets where deviations do occur, but the general behavior is reproduced and substantial $(x,Q^2)$ ranges are well fitted.
	\GI{
		The large $Q^2$-range covered by the HERA data in Fig.\ \ref{Fig: F2smallx} is quite well described as given by the DGLAP equations and can be further improved by switching to NNLO DGLAP evolution. Because we are also interested in polarized parton distributions \cite{Ekstedt:2018ddj} where splitting functions are only known to NLO precision, we choose to use NLO evolution for both the polarized and the unpolarized distributions.
	}
	
\begin{figure}
		\centering
		\includegraphics[width=1.0\columnwidth,trim=-0.0cm 2.0cm -0.0cm 0cm,clip]{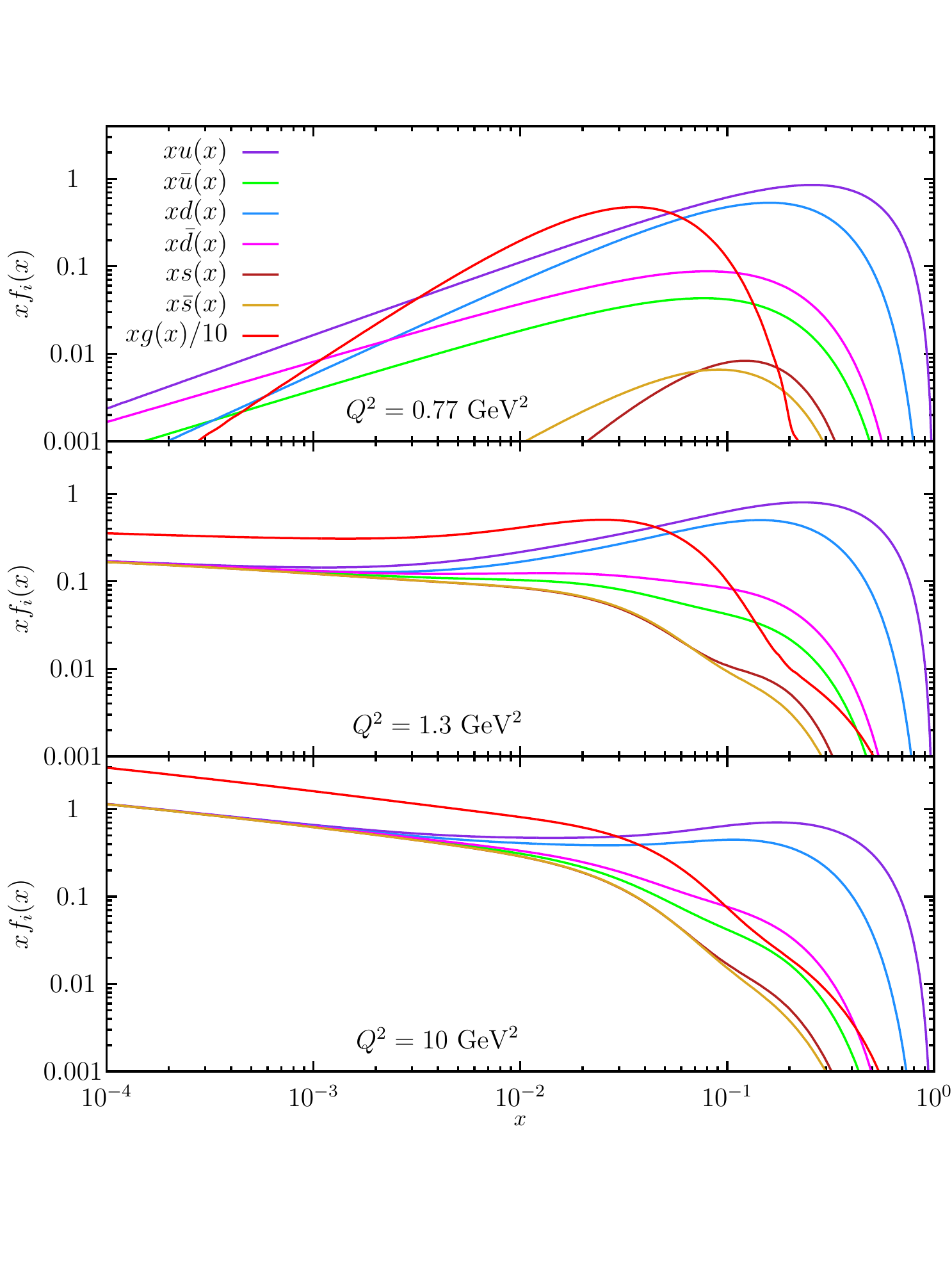}
		\caption{The resulting PDFs $xf_i(x)$ of the proton at the starting scale $Q_0^2 = 0.77~\text{GeV}^2$ and at $Q^2 = 1.3~\text{GeV}^2$ and	 $Q^2 =10~\text{GeV}^2$. }\label{Fig: protonPDFs}
	\end{figure}

	It is therefore of interest to look into some details on the $x$-shapes of individual parton densities as they emerge from the model including both the hadronic fluctuations and the probed hadron's generic parton density description, but without any pQCD evolution. This is shown in the top panel of Fig.\ \ref{Fig: protonPDFs}, where the overall shape of the valence quark distributions is quite similar to conventional PDF parametrizations. 
	\GI{
		The fact that the distributions go smoothly to zero, $xf_i(x)\to 0$, for $x\to 1$ is not due to a choice of a particular form of the PDFs, such as including a factor $(1-x)^a$ as in most parametrizations of PDFs. Instead this behavior is due to the kinematical constraints on the DIS process, as explained above in connection with Eq.~(\ref{E: f-bare-x}). This is particularly important for a proper model of the valence quark distributions. Further characteristic features are that the gluon distribution is quite large for smaller $x$ and the sea quarks are suppressed but not at all negligible.
	} 
	So there is a non-trivial contribution of non-perturbatively generated sea quarks in the bound-state proton. Examining the sea quark distributions one notes the different distributions for $\bar{u}$ and $\bar{d}$, on the one hand, and for $s$ and $\bar{s}$, on the other. This is the basis for asymmetries in the light sea and strange sea, as will be further discussed below. 
	
	The effect on the PDFs from pQCD evolution using the DGLAP equations is shown in the middle and lower panels of Fig.\ \ref{Fig: protonPDFs}. Due to the $\log{Q^2}$-dependent evolution there is a quick increase from $Q_0^2$ so that already at $Q^2=1.3$~GeV$^2$ the perturbatively generated sea quarks and gluons dominate at small $x$ over the originally non-perturbative sea. 
	
	The PDFs obtained at the starting scale $Q^2_0$ are evaluated numerically. However, for illustrative purposes the starting distributions for a parton $i$ can be parametrized in the convenient form $x f_i(x)=a~x^b (1-x)^c$.  The fitted coefficients for the various distributions are given in Table \ref{Table: xdistPars}.
	
	\begin{table}[h]\caption{Parametrization $xf_i(x)=ax^b(1-x)^c$ at $Q_0^2$. }\label{Table: xdistPars}
		\renewcommand{\arraystretch}{1.2}
		\begin{center}
			\begin{tabular}{|P{2.2cm}|P{1.2cm}|P{1.2cm}|P{1.2cm}|}
				\hline
				Distribution & $a$ & $b$ & $c$\\ \hline
				$x \bar{s}$ & 0.81 & 1.4 & 14.\\ \hline
				$x s$ & 6.7 & 2.2 &16.\\ \hline
				$x \bar{d}$ & 0.97 & 0.69 & 7.9 \\ \hline
				$x d$ &11. &1.1 & 5.8 \\ \hline
				$x \bar{u}$ & 0.54 &0.71 & 8.6 \\ \hline
				$x u$ & 5.5 & 0.84 & 2.4 \\ \hline
				$x g$ & 5700. & 1.6 & 47. \\ \hline
			\end{tabular}
		\end{center}
	\end{table}
	
	\subsection{The $\bar{d}$-$\bar{u}$ asymmetry}\label{Sec: ubardbarasymm}
\begin{figure}
		\centering
		\scalebox{1}{
			\includegraphics[width=1\columnwidth,trim=0cm 0cm 0cm 0cm,clip]{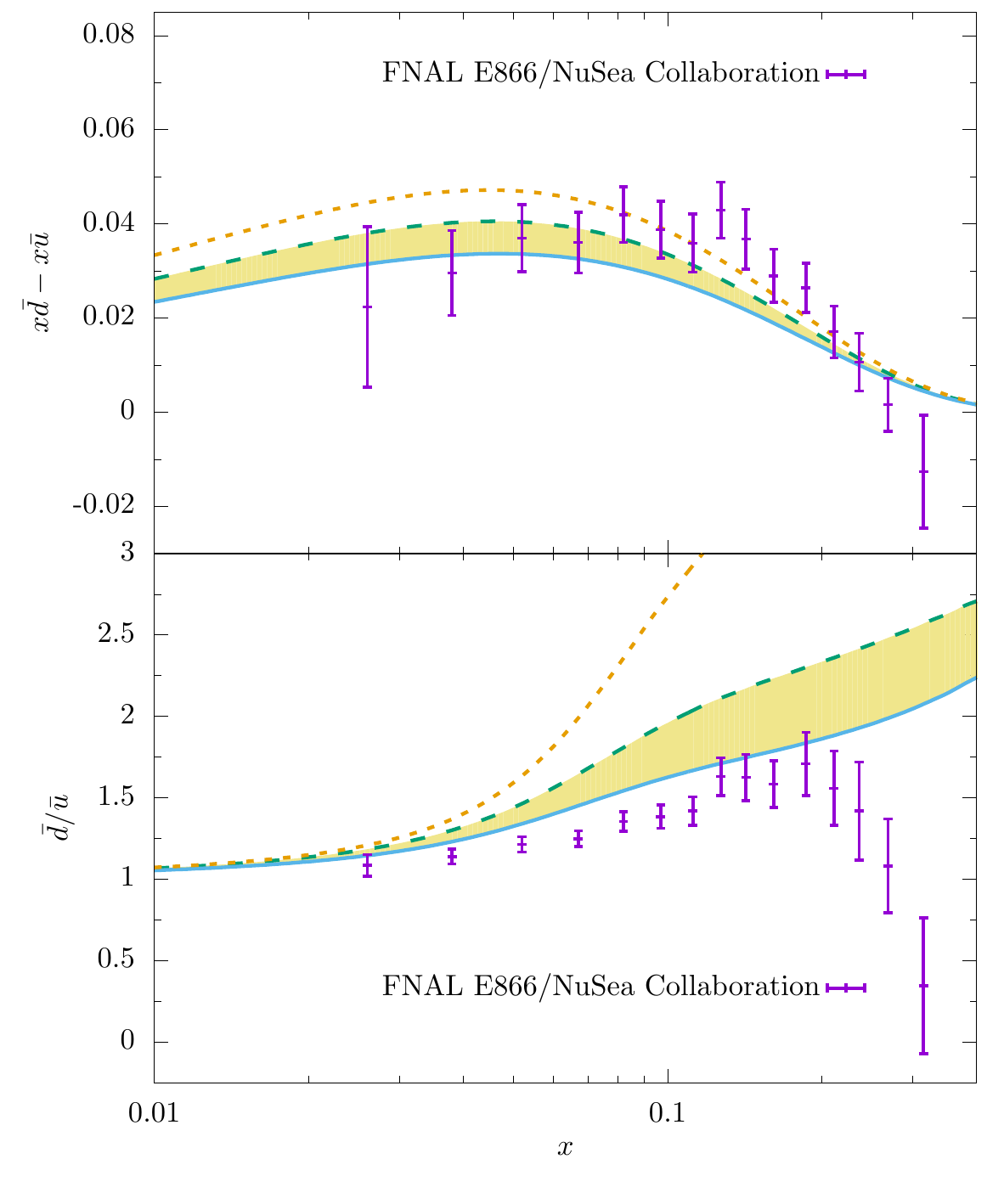}
		}
		\caption{The light-quark sea asymmetry in terms of the difference $x \bar{d}(x)-x\bar{u}(x)$ (upper panel) and the ratio $\bar{d}(x)/\bar{u}(x)$ (lower panel).  Data from Fermilab E866/NuSea collaboration \cite{Towell:2001nh} compared to our model result. The dotted curve takes into account only nucleon-pion ($N\pi$) fluctuations. The shaded band accounts for all fluctuations, with $\Delta\pi$ being most important among the decuplet contributions. The shaded band is obtained by a variation in the decuplet coupling $h_A/m_\Delta$ where the solid (dashed) curve refers to the largest (smallest) value of the decuplet coupling in Eq.\ (\ref{E: 1020}).}\label{Fig: VaryhA} 
	\end{figure}

	From a pQCD point of view, the momentum distribution of the $\bar{d}$ and $\bar{u}$ sea in the proton should be similar because $m_u, m_d\ll\Lambda_\text{QCD}$, $Q_0$. This is, however, not the case as seen in data from e.g.\ \cite{Towell:2001nh} where a clear asymmetry is seen (cf.\ Fig.\ \ref{Fig: VaryhA}). Such an asymmetry arises naturally from hadronic fluctuations of the proton where the non-perturbative sea distributions are dominantly generated by the pions \AN{\cite{Thomas:1983fh,Melnitchouk:1991ui,Wakamatsu:1991yu,Henley:1990kw,Carvalho:1999he,Holtmann:1996be,Kumano:1997cy, Speth:1996pz, Kofler:2017uzq}}. The energy-wise lowest fluctuations are $P\pi^0$ and $n\pi^+$, where the former does not contribute to the $\bar{d}$-$\bar{u}$ asymmetry Because the $\pi^0$ is symmetric in $\bar{d}d$ and $\bar{u}u$. Taking only these nucleonic fluctuations into account gives already decent agreement with data on the difference $x\bar{d}-x\bar{u}$ as shown by the dotted curve in Fig.\ \ref{Fig: VaryhA} (upper panel). However, these nucleonic fluctuations are not sufficient to explain the ratio $\bar{d}(x)/\bar{u}(x)$ as shown by the dotted curve in the lower panel of Fig.\ \ref{Fig: VaryhA}. 
	
	The results become better when also including fluctuations with other baryons. In particular the $\ket{\Delta^{++}\pi^-}$ state, having the largest decuplet coupling (see Table \ref{Table: Coupling}) and having $\bar{u}$ in the $\pi^-$, contributes significantly to bring the curves down to the data points. The full octet and decuplet contribution is shown in Fig.\ \ref{Fig: VaryhA} where the band represents a variation in the decuplet coupling $h_A/m_\Delta$, with the largest (smallest) value in Eq.\ (\ref{E: 1020}) corresponding to the solid (dashed) curve. As seen in the figure an $n\%$ variation in the coupling results in an $n\%$ variation in the difference $x\bar{d}-x\bar{u}$ for small $x\lesssim 0.15$. The variation has a slightly smaller impact on the ratio $\bar{d}/\bar{u}$, but the variation is essentially of the same order of magnitude as that of $x\bar{d}-x\bar{u}$. 
	
	\subsection{The strange sea of the proton}\label{Sec: StrangeSea}
\begin{figure}
		\centering
		\scalebox{.745}{
			\includegraphics{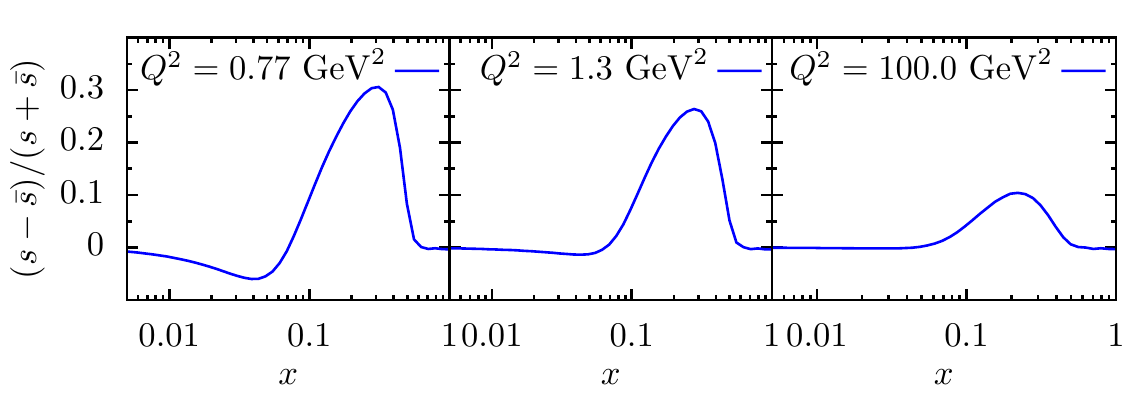}
		}
		\caption{The ratio $(s-\bar{s})/(s+\bar{s})$ as a function of $x$ evaluated for different values of $Q^2$. }
		\label{Fig: ssbarRatio} 
	\end{figure}

	Due to the possibility of the proton to fluctuate into $\ket{\Lambda K^+}$, $\ket{\Sigma K}$ and $\ket{\Sigma^* K}$ states a non-perturbative strange sea will arise, as shown in Fig.\ \ref{Fig: protonPDFs}. It is suppressed relative to the light-quark sea partly due to the kinematical suppression of these fluctuations with higher-mass hadrons, but also due to the smaller hadronic couplings shown in Table \ref{Table: Coupling}. Moreover, the $x$-distributions of $s$ and $\bar{s}$ are not the same, but $s$ has a harder momentum distribution than $\bar{s}$ \GI{\cite{Signal:1987gz,Melnitchouk:1996fj,Melnitchouk:1999mv,Alwall:2004rd}}. This is a kinematic effect arising from that the $s$ quark is in the baryon, due to its higher mass, gets a harder $y$-spectrum in the hadronic fluctuation than the lighter meson containing the $\bar{s}$. The dominance of kaons in the low-$x$ region and similarly the dominance of strange baryons in the higher-$x$ region is clearly seen in the ratio $(s-\bar{s})/(s+\bar{s})$ in Fig.\ \ref{Fig: ssbarRatio}. Here one can also see how the additional symmetric $s\bar{s}$ from $g\to s\bar{s}$ in pQCD reduces this ratio with increasing $Q^2$. Because pQCD fills up the low-$x$ region to a higher degree, the kaon effect is more depleted than the `baryon peak', which is, however, shifted to lower $x$.
	The symmetric $s\bar{s}$ sea from the $\log{Q^2}$ DGLAP evolution builds up quickly and dominates at small-$x$ already for $Q^2=1.3~\text{GeV}^2$, as shown in the middle panel of Fig.\ \ref{Fig: ssbarRatio}. Thus, the asymmetry is only expected to be visible at quite low $Q^2$ and therefore hard to observe experimentally. 
	
\begin{figure}
		\centering
		\scalebox{1}{
			\includegraphics[width=1\columnwidth]{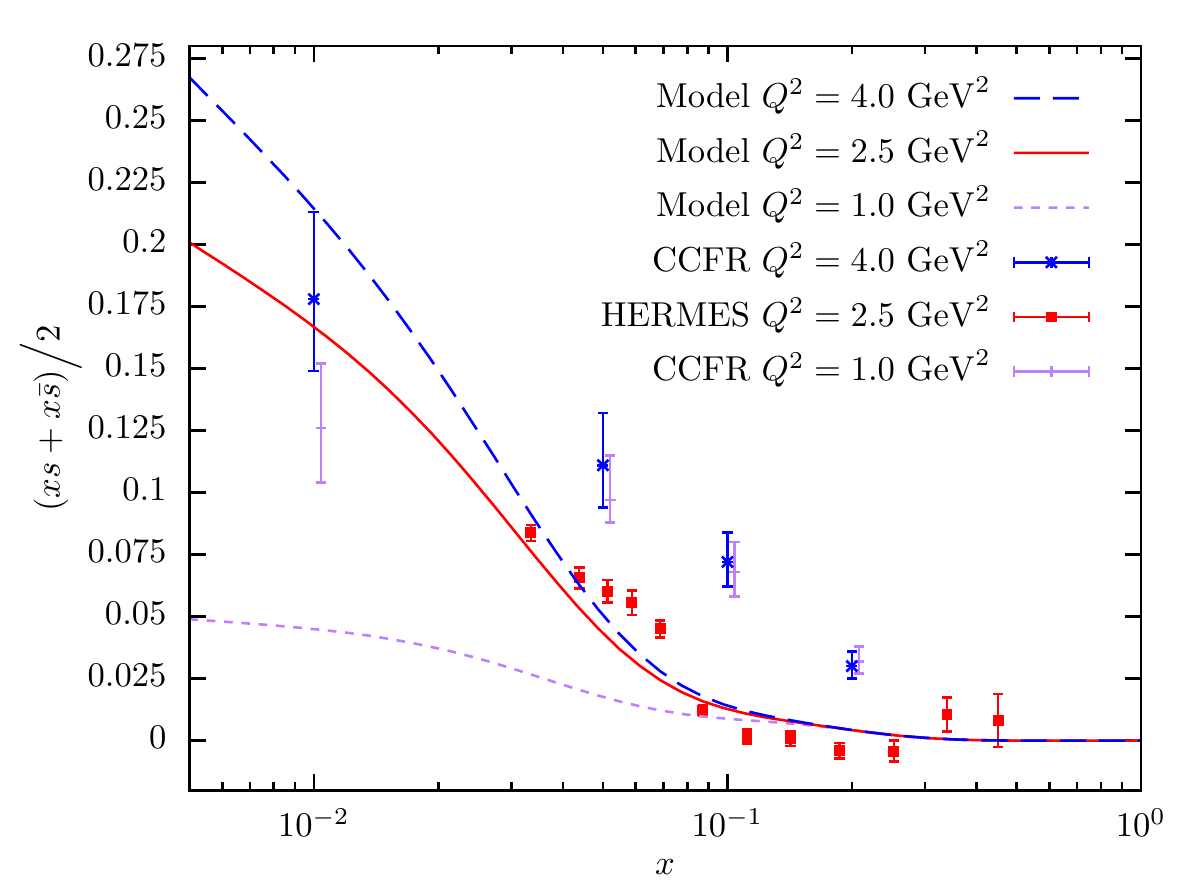}
		}
		\caption{The strange-quark sea, $(xs+x\bar{s})/2$, as a function of $x$ for different values of $Q^2$, with our model results compared to data from \cite{Bazarko:1994tt,Airapetian:2013zaw}. The CCFR analysis assumes $xs(x) = x\bar{s}(x)$. }
		\label{Fig: ssbar1} 
	\end{figure}

	The extraction of the strange sea from data is not at all trivial because it requires some additional observable  to signal that an $s$ or $\bar{s}$ has been probed. In Fig.~\ref{Fig: ssbar1} our model is compared to data on the total strange sea $\left(xs(x)+x\bar{s}(x) \right)/2$. The CCFR data \cite{Bazarko:1994tt} are obtained from neutrino-nucleon scattering producing a charm quark decaying semileptonically giving an opposite sign dimuon signature, i.e. 
	$\nu_\mu + N\to \mu^- + c + X$ where $c\to s + \mu^+ + \nu_\mu$ or
	$\bar{\nu}_\mu + N\to \mu^+ + \bar{c} + X$ where $\bar{c}\to \bar{s} + \mu^- + \bar{\nu}_\mu$.  
	The charged-current subprocess $W^+s\to c$ or $W^- \bar{s}\to \bar{c}$ is here the essential point. 
	Other sources of charm production, such as $W^+g\to c\bar{s}$ or $W^-g\to \bar{c}s$, or other sources of dimuon production from other decays must be taken into account to extract a proper measure of the strange sea, as discussed in \cite{Bazarko:1994tt}. The result shows that although the shape difference between the $xs(x)$ and $x\bar{s}(x)$ distributions is consistent with zero, it has large uncertainties. CCFR assumed $xs(x)=x\bar{s}(x)$ for extracting the data points shown in Fig.\ \ref{Fig: ssbar1}.
	
	The more recent result of HERMES \cite{Airapetian:2013zaw} is obtained from data on the multiplicities of charged kaons in semi-inclusive deep-inelastic electron-proton scattering. This requires a detailed and non-trivial analysis of the fragmentation function into kaons to extract the contribution from initial-state strange quarks in the basic DIS process $\gamma s\to s$ or $\gamma \bar{s}\to \bar{s}$. As seen in Fig.~\ref{Fig: ssbar1} the CCFR and HERMES results differ substantially and do not provide a clear result on the strange sea. Our model result agrees reasonably well with the HERMES result, but compared to CCFR it has a too small strange sea at low $Q^2$. Because the strange-quark sea is not yet well determined, we contribute with some further investigations.

\begin{figure}
		\centering
		\scalebox{.97}{
			\includegraphics[width=1\columnwidth,trim=0.20cm 0.0cm 0cm 0cm,clip]{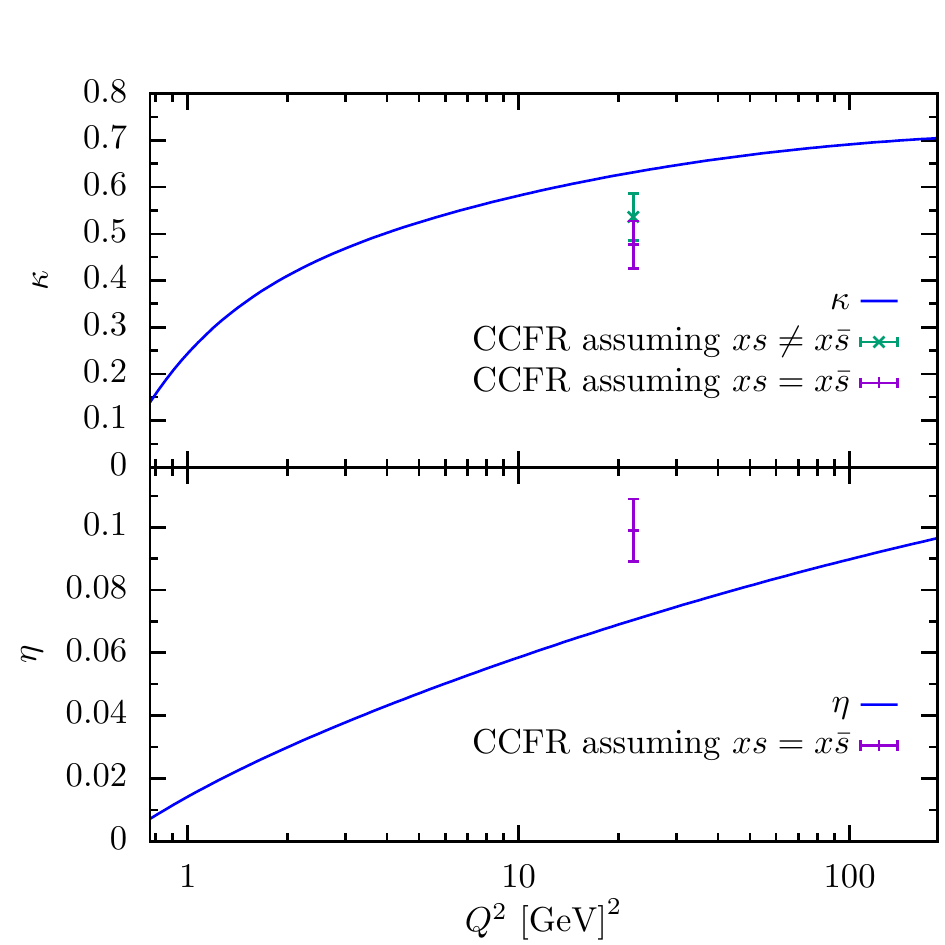}
		}
		\caption{
			The ratios $\kappa$ (upper panel) and $\eta$ (lower panel) in Eq.\ (\ref{E: Kappa-Eta}) for the strange-quark content of the proton as calculated in our model. The data points at $Q^2=22.2~ \text{GeV}^2$ are from the CCFR Collaboration \cite{Bazarko:1994tt}.}\label{Fig: kappaeta} 
	\end{figure}

	The strange-quark content of the proton can be characterized by the momentum fraction carried by the strange sea relative to the light-quark sea or the non-strange quark content \cite{Bazarko:1994tt}
	\eq{\label{E: Kappa-Eta} 
		&\kappa = \frac{\int_0^1\!\dd x\, [xs\left(x, Q^2\right)+x\bar{s}\left(x,Q^2\right)]}{\int_0^1\!\dd x\,
			[
			x\bar{u}\left(x, Q^2\right)+x\bar{d}\left(x,Q^2\right)
			] },
		\\&\eta = \frac{\int_0^1\!\dd x\,[xs\left(x, Q^2\right)+x\bar{s}\left(x,Q^2\right)] }{\int_0^1\!\dd x\,
			[
			x{u}\left(x, Q^2\right)+x{d}\left(x,Q^2\right)
			]
		},
	}
	where $\kappa=1$ would mean a flavor SU(3) symmetric sea. These ratios are shown in Fig.\ \ref{Fig: kappaeta} versus $Q^2$, where the qualitative behavior is understandable within our model. At $Q_0^2$ there is, as discussed, only a small non-perturbative strange-quark sea from hadron fluctuations. With increasing $Q^2$ the perturbative $\log{Q^2}$ evolution first builds up the $s\bar{s}$ sea quickly and then flattens off at larger scales (note the logarithmic $Q^2$ scale in the figure). 
	
\begin{figure}[t]
		\centering
		\scalebox{0.74}{
			\includegraphics{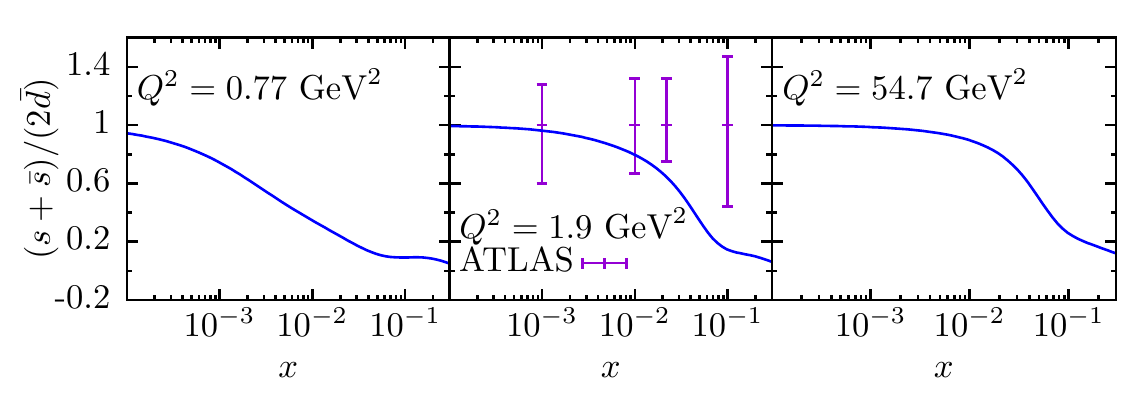}
		}
		\caption{The strange-sea suppression factor $r_s = (s+\bar{s})/(2\bar{d})$ as a function of $x$ evaluated for different values of $Q^2$. Data from the ATLAS collaboration \cite{Aad:2012sb, Aad:2014xca}. } 
		\label{Fig: suppressionssbar} %
	\end{figure}

	The proton sea is, however, not flavor SU(3) symmetric as indicated by the value of $\kappa$ and quantified by the strange-sea suppression factor
	\eq{
		r_s(x,Q^2) = \frac{s(x,Q^2)+\bar{s}(x,Q^2)}{2\bar{d}(x,Q^2)}. 
	} 
	Our results for this quantity are shown in Fig.\ \ref{Fig: suppressionssbar} together with ATLAS data \cite{Aad:2012sb, Aad:2014xca}. As seen for $Q^2$ slightly larger than the starting value for the QCD evolution $Q_0^2$, the suppression factor is constant and near unity for $x\lesssim 0.01$. For low $x$ this is in agreement with the $ePWZ$-fit of \cite{Aad:2012sb}. For larger $x$ our model gives $r_s\left(0.023,1.9~\text{GeV}^2\right) \approx 0.62$, which is consistent within uncertainties of experimental observations: $r_{s}(0.023,1.9~\text{GeV}^2) = 0.56 \pm 0.04$ \cite{Alekhin:2014sya}, $r_s(0.023,1.9~\text{GeV}^2) = 1.00^{+0.25}_{-0.28}$ \cite{Aad:2012sb} and $r_s(0.023,1.9~\text{GeV}^2) = 0.96^{+0.26}_{-0.30}$ \cite{Aad:2014xca}. As seen in Fig.\ \ref{Fig: suppressionssbar}, $r_s\rightarrow 1 $ as $x\rightarrow 0$, this supports the hypothesis that the quark sea at low $x$ is flavor symmetric.

	\begin{figure}[t]
		\centering
		\scalebox{0.92}{
			\includegraphics{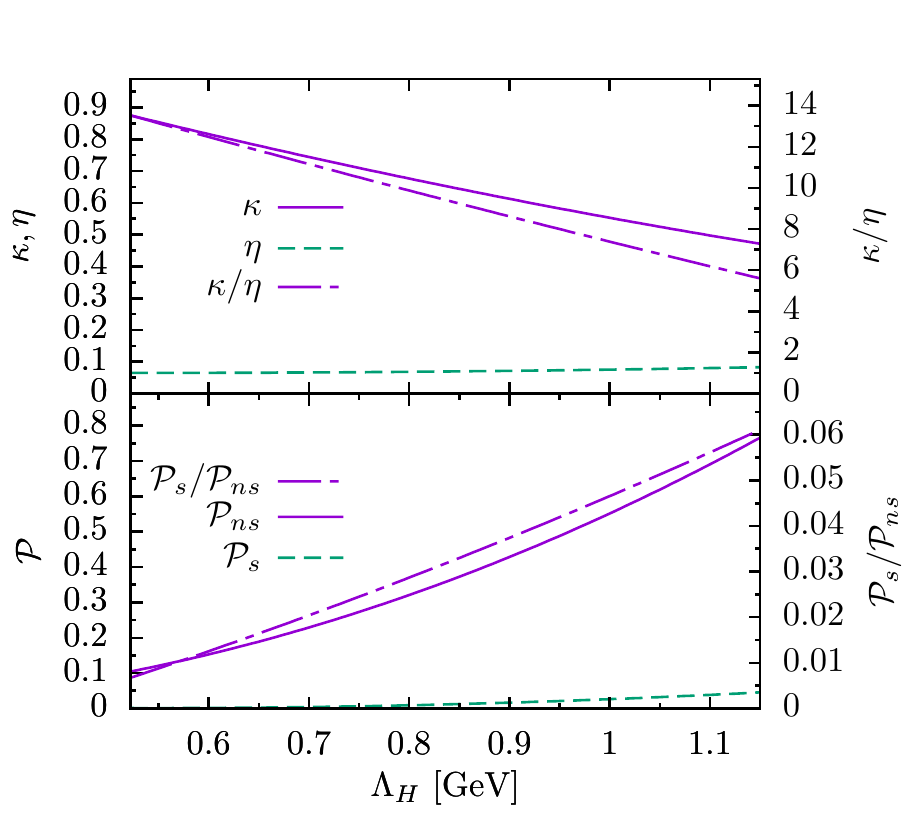}
		}
		\caption{Upper panel: The ratios $\kappa$ and $\eta$, evaluated at $Q^2=22.2~\text{GeV}^2$, as function of the hadron fluctuation regulator $\Lambda_H$. The middle curve is the ratio $\kappa/\eta$.\ Lower panel: The probability for all the hadronic fluctuations containing strangeness (dashed curve) and not containing any strangeness (solid curve), as a function of $\Lambda_H$. The ratio of these probabilities is also shown (dot-dashed curve). Notice the two different scales on the vertical axes. }
		\label{Fig: ketaLambdaSnonSratio} %
	\end{figure}
	
	
	For completeness we show in Fig.\ \ref{Fig: ketaLambdaSnonSratio} (top panel) the dependence of the strange-sea ratios $\kappa$ and $\eta$ on the hadron fluctuation regulator $\Lambda_H$. Whereas $\kappa$ strongly depends on $\Lambda_H$, $\eta$ is almost independent of $\Lambda_H$. This can be understood from the plot in the lower panel of the same figure which compares the non-strange fluctuation probability $\mathcal{P}_{ns}$ (e.g.\ for $\ket{N\pi}$ and $\ket{\Delta\pi}$) and the probability $\mathcal{P}_s$ that the proton fluctuates into a hadron pair that does contain strangeness. Not only is $\mathcal{P}_s\ll \mathcal{P}_{ns}$, but also its slope is much smaller implying that for increasing $\Lambda_H$ the rate of population growth is much larger for those fluctuations that contain $\bar{u}$ and $\bar{d}$ quarks, than those containing $s$ and $\bar{s}$ quarks ($\Delta \mathcal{P}_s/ \Delta \Lambda_H \approx 5\% ~\text{GeV}^{-1}$ and $\Delta \mathcal{P}_{ns}/ \Delta \Lambda_H \approx 90\% ~\text{GeV}^{-1}$ between $0.5~\text{GeV}\leq \Lambda_H \leq 1.0~\text{GeV}$). Hence $\kappa$ depends much more strongly on $\Lambda_H$ than does $\eta$ due to appearance of $\bar{u}$ and $\bar{d}$ distributions in its definition. As shown in the lower panel of Fig.\ \ref{Fig: ketaLambdaSnonSratio}, at the regulator value of $\Lambda_H=0.87$ GeV, roughly 1\% of the fluctuations contain strangeness. This can be compared to the result obtained in Ref.\ \cite{Alwall:2005xd}, where the strangeness fluctuations had to constitute 5\% in order to reproduce the then available CCFR data.

	If it turns out to be a need for a larger non-perturbative strange-quark sea than in our present model, this might be remedied by a minor modification of the model. One option could be 
	a flavor-dependent momentum cutoff $\Lambda_H$, but to keep our model as simple as possible we refrained from introducing more parameters. An alternative explanation might come from the importance of additional degrees of freedom not considered so far. In the strangeness $S=-1$ sector there are four baryonic states below the antikaon-nucleon threshold: $\Lambda$, $\Sigma$,
	$\Sigma^*(1385)$ and $\Lambda^*(1405)$. The first three have been taken into account in our approach as the strangeness
	counterparts of the nucleon and the $\Delta(1232)$ considered in the pion-baryon fluctuations. But we have not included the
	$\Lambda^*(1405)$ in our framework. On the one hand, we found that the comparatively heavy $K\Sigma^*(1385)$ fluctuation is much less important than the lighter $K \Lambda$. This suggests that also $K\Lambda^*(1405)$ is negligible. On the other hand, the negative-parity $\Lambda^*(1405)$ couples with an $s$-wave to nucleon-antikaon while all our interactions are of $p$-wave nature. This can enhance the importance of the $\Lambda^*(1405)$. The ultimate reason why we have not explored its influence in the present work is the absence of unambiguous experimental information about the coupling strength between a nucleon and $K\Lambda^*(1405)$. This is related to the long-standing question about the nature of the $\Lambda^*(1405)$. Being lighter than all non-strange baryons with negative parity, it has been speculated because a long time \cite{Dalitz:1967fp} that the $\Lambda^*(1405)$ is merely an antikaon-nucleon bound state instead of a three-quark state; see, for instance \cite{Hall:2014uca} for further discussion and references. This would point to a relatively large coupling strength. Yet in view of these theoretical uncertainties we have not pursued a detailed analysis of the $K\Lambda^*(1405)$ fluctuation as long as there is no clear need for an enhancement of the strange sea.
	
	\subsection{Pion PDF}\label{Sec: Pion}
	\GI{
		With our model parameters fixed by DIS data, it is possible to find the distributions for other hadrons. The pion PDFs can be obtained by the method given in Section \ref{Sec: Distribution}. Here we do not take into account hadronic fluctuations for the following reasons. From the point of view of many-body theory the first contributions that one would consider are hadronic two-body fluctuations. However, due to parity conservation the pion cannot fluctuate into a pair of pions. Two-body states with a pion and a mesonic resonance are already quite far away from the pion mass shell. This applies even more to nucleon-antinucleon fluctuations. Next one might consider three-body fluctuations, in particular three pions. But for the nucleon we have not taken into account three-body states like two pions and a nucleon. There one might even run into a double-counting problem because the $\Delta$ is essentially an elastic resonance in the pion-nucleon channel, i.e.\ pion-$\Delta$ fluctuations contain a significant part of the three-body fluctuations into two pions and a nucleon. These reservations apply also to the pion. In addition, the three-pion fluctuations are suppressed at low energies due to the chiral dynamics of the pions. In technical terms, the derivative couplings of the Goldstone bosons lead to a suppression of the two-loop diagrams \cite{Scherer:2002tk} that correspond to the three-pion fluctuations. For all these reasons we will neglect hadronic effects for the pion.
	} 
	
	\begin{figure}[t]
		\includegraphics[width=0.5\textwidth]{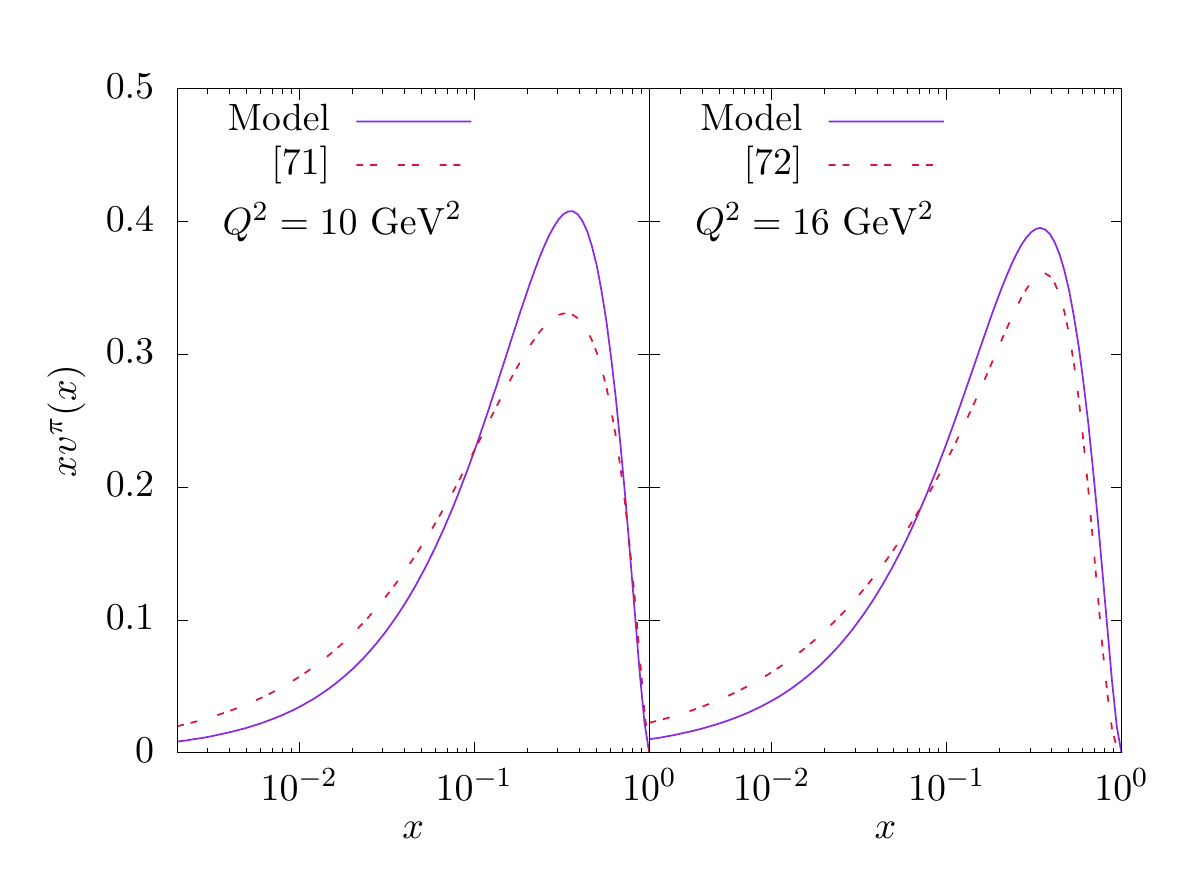}
		\caption{\GI{Model predictions for the valence quark distribution $x v^{\pi}=x u^{\pi^{+}}-x\bar{u}^{\pi^{+}}$, compared with parametrizations from \cite{Barry:2018ort} (left) and \cite{Aicher:2010cb}} (right). }\label{Fig: pionPDFs}
	\end{figure}
	
	\GI{
		Because experimental data are limited to the pion valence region, we choose to focus on this region. The comparison of our model with parametrizations from other groups \cite{Barry:2018ort,Aicher:2010cb} is shown in Fig.\ \ref{Fig: pionPDFs}. Obviously, we obtain a quite good agreement, which gives further credits to our generic model for the parton distributions in a hadron. It would be interesting to compare our predictions for other mesons with data. However, because our primary interest lies in the nucleon we leave this comparison for a dedicated study.
	}
	
	\section{Conclusions}\label{Sec: Conclusions}
	
	This study has demonstrated that the momentum distribution of partons in the proton, and thereby the observed proton structure functions, can be understood in terms of basic physical processes. We thereby obtain new knowledge regarding the poorly understood non-perturbative dynamics of the bound-state proton. Using the well-established pQCD DGLAP equations for the $Q^2$-dependence above the scale $Q_0^2$, our model developed here addresses the basic shape in the distribution of the energy-momentum fraction $x$ carried by different parton species at $Q_0^2$. Thus, the model treats the physics at the transition from bound-state hadron degrees of freedom to the internal parton degrees of freedom. It does so by convoluting hadronic quantum fluctuations with partonic fluctuations. To describe the former we use the leading-order Lagrangian of chiral perturbation theory, the low-energy effective theory that respects the symmetries of QCD as the underlying theory. The partonic fluctuations arise quantum mechanically due to confinement within the small size of a hadron, as given by the uncertainty relation in position and momentum. 
	
	Interestingly the fit that gives best agreement with the structure functions $F_2$ and $xF_3$ data yields a value where the hadronic language ends and QCD evolution begins to be $\Lambda_H =  Q_0=0.87~\text{GeV}$.

	Thus, having a model with effectively only four dimensionful parameters with physically meaningful values, it is highly non-trivial that we obtain a very satisfying reproduction of a large amount of data. 
	\GI{
		The model's assumption on the origin of non-perturbative sea quarks at the scale $Q_0^2$ being only given by the hadronic fluctuations, works well to describe data when also including the perturbatively generated sea quarks by the DGLAP evolution at higher $Q^2$. Thus, there is no need to complicate the model with some additional, unknown source of a non-perturbative quark sea. 
	}In particular we find that the $n\pi^+$ and the $\Delta\pi$ fluctuations generate the flavor asymmetry $x\bar{d}-x\bar{u}>0$ in the proton sea, to a large extent consistent with experimental data. This shows that the model captures the essential physics observed.
	
	\GI{
		Regarding the strange-quark sea of the proton, arising from proton fluctuations into strange hadrons, we find that it is substantially suppressed due to the larger masses of strange hadrons. 
	}An asymmetry in terms of different $x$-distribution for $s$ and $\bar{s}$, with $s(x)$ being harder, is found. However, this effect is reduced at larger $Q^2$ due to the development of the symmetric $s\bar{s}$ sea from $g\to s\bar{s}$ in pQCD. The remaining asymmetry at observed $Q^2$ is too small to be seen in present data. Further details of the non-perturbative strange sea of our model are given to promote future studies, including the potentially interesting inclusion of the $\Lambda^*(1405)$ in the hadronic fluctuations.
	
	We have here considered PDFs of the proton and \GI{the pion} where most experimental information is available for testing our model. The model is, however, quite general and can give the parton momentum distributions in any hadron.
	
	Based on the phenomenological success of the model and its theoretical basis, spin degrees of freedom and the proton spin puzzle are studied in another paper \cite{Ekstedt:2018ddj}.

	\section*{Acknowledgments}
	We acknowledge helpful discussions with C.\ G.\ Granados and U.\ Aydemir at an early stage of this project. This work was supported by the Swedish Research Council under contract 621-2011-5107. 
	
	\appendix

	\section{The relevant Lagrangians}\label{App: Lagrangian}
	For the metric we use $g= \text{diag}(+1,-1,-1,-1)$ and $\epsilon^{0123}=+1$. 
	The relevant part of the leading-order chiral Lagrangian describing the interaction of Goldstone bosons with nucleons and spin 3/2 baryons is given by \cite{Jenkins:1991es, Pascalutsa:1999zz, Pascalutsa:2006up, Ledwig:2014rfa} 
	\eq{\label{E: DNpiAppendix}
		\mathcal{L}_\text{int} =~&\frac{D}{2}\tr(\bar{B} 
		\gamma^\mu \gamma_5 \{u_\mu,B\})
		+\frac{F}{2}\tr (\bar{B}\gamma^\mu\gamma_5[u_\mu,B])
		\\
		+&\frac{1}{2\sqrt{2}}h_A\epsilon_{ade}g_{\mu\nu}\left(\bar{T}^\mu_{abc}u^\nu_{bd} B_{ce}+\bar{B}_{ec}u^\nu_{db} T_{abc}^\mu\right). 
	}
	Relativistic Rarita-Schwinger fields exhibit some problematic features related to how to handle its spin-1/2 components. Apart from exchanging spin-3/2 resonances, the Lagrangian (\ref{E: DNpiAppendix}) induces an additional unphysical contact interaction. This can be cured by subscribing to the Pascalutsa prescription, which in our case means making the substitution \cite{Pascalutsa:1999zz, Pascalutsa:2006up}
	\eq{\label{E: Pascalutsa}
		T^\mu\rightarrow -\frac{1}{m_R}\epsilon^{\nu\mu\alpha\beta}\gamma_5\gamma_\nu \partial_\alpha T_\beta, 
	}
	where $m_R$ refers to the resonance mass ($m_R = m_\Delta, m_{\Sigma^*}$). 
	Note that this substitution induces an explicit flavor breaking but these effects are beyond leading order. 
	
	In (\ref{E: DNpiAppendix}) $B_{ab}$ is the entry in the $a$th row, $b$th column of the matrix representing the octet baryons  
	\eq{
		B = 
		\begin{pmatrix}
			\frac{1}{\sqrt{2}}\Sigma^0+\frac{1}{\sqrt{6}}\Lambda       & \Sigma^+ & P \\
			\Sigma^-       & -\frac{1}{\sqrt{2}}\Sigma^0+\frac{1}{\sqrt{6}}\Lambda & n \\
			\Xi^-       & \Xi^0 & -\frac{2}{\sqrt{6}}\Lambda
		\end{pmatrix}. 
	}
	The Goldstone bosons are contained in 
	\eq{
		\Phi = 
		\begin{pmatrix}
			\pi^0+\frac{1}{\sqrt{3}}\eta      & \sqrt{2}\pi^+ & \sqrt{2}K^+ \\
			\sqrt{2}\pi^-       & -\pi^0+\frac{1}{\sqrt{3}}\eta& \sqrt{2}K^0 \\
			\sqrt{2}K^-       & \sqrt{2}\bar{K}^0 & -\frac{2}{\sqrt{3}}\eta
		\end{pmatrix} 
	}
	and $u_\mu$ is essentially given by 
	\eq{
		u_\mu \equiv \ii u^\dagger(\partial_\mu U)u^\dagger  = u_\mu^\dagger~\text{where}~u^2 \equiv U = \exp(\ii \Phi /F_\pi).
	}
	%
	Finally, the decuplet is represented by a totally symmetric flavor tensor 
	\eq{
		&T^{111} = \Delta^{++},T^{112} = \frac{1}{\sqrt{3}}\Delta^{+},T^{122} = \frac{1}{\sqrt{3}}\Delta^{0}, 
		T^{222} = \Delta^{-} \! ,
		\\
		& 
		T^{113} = \frac{1}{\sqrt{3}}\Sigma^{*+}, T^{123} = \frac{1}{\sqrt{6}}\Sigma^{*0}, 
		T^{223} = \frac{1}{\sqrt{3}}\Sigma^{*-},
		\\
		& 
		T^{133} = \frac{1}{\sqrt{3}}\Xi^{*0}, 
		T^{233} = \frac{1}{\sqrt{3}}\Xi^{*-}, 
		T^{333} = \Omega. 
	}
	
	The couplings we use \cite{Granados:2017cib} are $F_\pi = 92.4\text{ MeV},D=0.80,F=0.46$ \cite{Cabibbo:2003cu} and $h_A$ can be determined from the partial decay width $\Sigma^{*}\rightarrow \Lambda\pi$ or from $\Delta\rightarrow N\pi $ to be 
	\eq{
		h_A^{\Sigma^{*}\rightarrow \Lambda\pi} = 2.4  ~\text{and}~ 
		h_A^{\Delta\rightarrow N\pi} = 2.88. 
	}
	In the large-$N_C$ limit \cite{Dashen:1993as,Pascalutsa:2005nd}, one also gets ($N_C=$ number of colors) 
	\eq{
		h_A^\text{large-$N_C$} = \frac{3}{\sqrt{2}}g_A = 2.67, 
	}
	where $g_A = F+D=1.26$. 
	We will explore the range
	\eq{\label{E: hApm}
		h_{A}^{\pm}= 2.7\pm 0.3.
	} 
	Notice that after the substitution (\ref{E: Pascalutsa}) it is the ratio $h_A/m_R$ that appears with each decuplet-baryon--meson term [cf.\ Eq.\ (\ref{E: DNpi})]. That is, on the probability level one has schematically 

	\eq{\label{E: varyhAvaryLambda}
		&\mathcal{P}(\Lambda_H) \sim g_{OM}^2T_\text{oct}(\Lambda_H)
		\\
		&+ \left(\frac{h_A}{m_\Delta}\right)^2 T_\Delta(\Lambda_H) 
		+
		\left(\frac{h_A}{m_{\Sigma^*}}\right)^2 T_{\Sigma^*}(\Lambda_H) +\cdots 
	}

	so that one could vary $h_A$ by $\sim10\%$ for each of the separate decuplet terms keeping the masses as shown but numerically it makes not much of a difference to instead use $m_R = m_\Delta$ for both terms and vary the ratio between its smallest and largest values $1.737 ~\text{GeV}^{-1} < h_A/m_\Delta < 2.435 ~\text{GeV}^{-1}$ coming from [cf.\ (\ref{E: hApm})] 

	\eq{
		\frac{h_A^-}{m_{\Sigma^*}} 
		& = 1.737 \, \mbox{GeV}^{-1}
		< \frac{h_A}{m_{\Delta}}  
		< \frac{h_A^+}{m_{\Delta}}
		= 2.435 \, \mbox{GeV}^{-1}. 
	}

	The effect of this variation is shown in Fig.\ \ref{Fig: VaryhA} and its effects on the probabilities are studied in Ref.\ \cite{Ghaderi:2018zei}.
	\section{The vertex functions}\label{Sec: VertexFunctions}
	The functions $S^{\lambda}\left(y,\vect{k}_\perp\right)$ are the amplitudes for a particular hadronic fluctuation of a proton with positive helicity and we calculate them using (on-shell) light-front spinors and the Lagrangian of Eq.\ (\ref{E: DNpi}). These amplitudes were calculated for a wide variety of hadronic fluctuations in \cite{Holtmann:1996be}. Our results are basically the same with minor differences due to a different choice of Lagrangian. Apart from different normalizations, our results for the functions $S^\lambda(y,\vect{k}_\perp)$ agree with those found in \cite{Pasquini:2006dv}. 
	We now present the vertex functions for both choices of the meson's `derivative momentum'.
	
	For meson momentum choice (\ref{E: pionmomA}) the amplitudes are
	\eq{\label{E: VonehalfA}
		&S^{+\frac{1}{2}}(y,\vect{k}_\perp)=-
		\frac{(m_B+m_P)(y m_P-m_B)}{\sqrt{y}}
		,
		\\&S^{-\frac{1}{2}}(y,\vect{k}_\perp)
		=-\frac{(m_B+m_P)k_\perp \ee^{\ii \phi}}
		{\sqrt{y}},
	} 
	where the transverse momentum $k_\perp=|\vect{k}_\perp| = |(k_1,k_2)|$ and $\phi\in[0,2\pi[$ is the angle in the polar form of the object $k_1+\ii k_2 = k_\perp\ee^{\ii \phi}$. 
	For meson momentum choice (\ref{E: pionmomB}) the amplitudes are
	\eq{
		\label{E: VonehalfB}
		&
		S^{+\frac{1}{2}}(y,\vect{k}_\perp)
		=-\frac{k_\perp^2+m_M^2 y-(1-y)^2 m_B m_P}
		{(1-y)\sqrt{y}}, 
		\\
		&
		S^{-\frac{1}{2}}(y,\vect{k}_\perp)=-\frac{(m_B+m_P)k_\perp \ee^{\ii 
				\phi}}{\sqrt{y}}.
	}
	
	For the case of a fluctuation into a spin-$3/2$ baryon and a spinless meson we find that the different helicity configurations yield the following amplitudes. For meson momentum choice (\ref{E: pionmomA}) the amplitudes are
	\begin{widetext}
		\eq{
			\label{E: VthreehalfA}
			&S^{+\frac{3}{2}}(y,\vect{k}_\perp)=\frac{\ii\ee^{-\ii\phi} m_B \left(m_B+y m_P\right)}{\sqrt{2 y} y}k_\perp ,
			\\&S^{+\frac{1}{2}}(y,\vect{k}_\perp)=\frac{-\ii}{\sqrt{6y}y} \left[k_\perp^2(2m_B+y m_P)+(y m_P+m_B)^2(ym_P-m_B)\right],
			\\&S^{-\frac{1}{2}}(y,\vect{k}_\perp)=\frac{\ii\ee^{\ii\phi }}{\sqrt{6y}y}
			\left[k_\perp^2 +(m_B+y m_P)(y m_P-2m_B) \right] k_\perp,
			\\&S^{-\frac{3}{2}}(y,\vect{k}_\perp)=\ii \frac{\ee^{2\ii \phi}m_B}{\sqrt{2y}y}k_\perp^2.
		}
		For meson momentum choice (\ref{E: pionmomB}) the amplitudes are
		\eq{\label{E: S32ChoiceB}
			&S^{+\frac{3}{2}}(y,\vect{k}_\perp)=\frac{\ii\ee^{-\ii\phi} m_B \left(m_B+y m_P\right)}{\sqrt{2 y} y}k_\perp ,
			\\
			&S^{+\frac{1}{2}}(y,\vect{k}_\perp) = 
			\mathrm{i}\frac{k_\perp^2 (m_P y-m_B (y-2))-\left(m_B^2 (y-1)^2-m_M^2
				y^2\right) (m_B+m_P y)}{\sqrt{6} (y-1) y^{3/2}},
			\\
			&S^{-\frac{1}{2}}(y,\vect{k}_\perp)=-\ii \frac{k_\perp 
				\ee^{\ii\phi}  \left[k_\perp^2+(y-1) m_B \left(y m_P-(y-2) 
				m_B\right)+m_M^2 y^2\right]}{\sqrt{6} (y-1) y^{3/2}},
			\\
			&S^{-\frac{3}{2}}(y,\vect{k}_\perp)
			=\ii \frac{\ee^{2\ii \phi}m_B}{\sqrt{2y}y}k_\perp^2.
		}
	\end{widetext}
	
	Notice that as $k_\perp$ goes to zero only the spin-conserving amplitudes $S^{+\frac{1}{2}}(y,\vect{k}_\perp)$ remain in (\ref{E: VonehalfA}-\ref{E: S32ChoiceB}), as expected. 
	
	We also note that in the chiral limit $m_M\to 0 $ with $y\to 1$ and $k_\perp\to 0$ the amplitudes corresponding to momentum choice (\ref{E: pionmomB}) vanish, in line with the Goldstone theorem \cite{PhysRev.127.965}. 
	
	
	\bibliographystyle{apsrev4-1}
	\bibliography{asymm_arxiv_v2.bib} 
\end{document}